\documentclass[aps,prd,superscriptaddress,nofootinbib,longbibliography,11pt]{revtex4-2}
\usepackage{amsmath,amssymb}
\usepackage{epsfig,graphicx}
\usepackage[font=small]{subcaption}
\usepackage{xcolor}
\usepackage{multirow}

\catcode`@=11
\@addtoreset{equation}{section}
\catcode`@=12

\usepackage{hyperref}
\hypersetup{
    colorlinks,%
    citecolor=blue,%
    filecolor=blue,%
    linkcolor=blue,%
    urlcolor=purple
}

\begin{document}

\title{\bf Deep learning in bifurcations of particle trajectories}
\author{Morteza Mohseni}
\email{m-mohseni@pnu.ac.ir}
\affiliation{Physics Department, Payame Noor University, Tehran 19395-3697, Iran}


\begin{abstract}
    We show that deep learning algorithms can be deployed to study bifurcations of particle trajectories. We demonstrate this for two physical
    systems, the unperturbed Duffing equation and charged particles in magnetic reversal by using the AI Poincar\'{e} algorithm. 
    We solve the equations of motion by using a fourth-order Runge-Kutta method to generate a dataset for each system. We use 
    a deep neural network to train the data. A noise characterized by a noise scale $L$ is added to data during the training. 
    By using a principal component analysis, we compute the explained variance ratios for these systems which depend on the noise scale. By 
    plotting explained ratios against the noise scale, we show that they change at bifurcations. For different values of the Duffing equation 
    parameters, these changes are of the form of different patterns of growth-decline of the explained ratios. For the magnetic reversal, the 
    changes are of the form of a change in the number of principal components. We comment on the use of this technique for other dynamical 
    systems with bifurcations. 
\end{abstract}

\maketitle

\section{Introduction}\label{intro}
Artificial intelligence (AI) and its various ramifications like machine learning or deep learning have been applied in almost all branches of 
science and technology and are affecting everyday life profoundly. AI-inspired techniques are being applied to problems in different branches
of physics increasingly; see \cite{doi:10.1146/annurev-conmatphys-031119-050745, 2018PhPl...25h0901S, 2019PhR...810....1M, 2022RvMP...94c1003B,
2021arXiv211200851A, RevModPhys.91.045002, DECELLE2022128154} for physics-oriented reviews. {For a more recent review, see
\cite{Wang2023}. For an example in the context of nonlinear physics, see \cite{Aksoy_2023}, and for a general discussion on the 
importance of AI in complex systems, see \cite{Boccaletti_2020}}. Rapid developments of new powerful AI algorithms 
seem to offer a promising direction to tackle widespread problems in physics.

In an interesting work \cite{2021PhRvL.126r0604L}, a deep learning algorithm, the so called AI Poincar\'{e}, has been developed by which it is 
possible to detect the conserved quantities of some Hamiltonian systems. The algorithm was tested there for five simple dynamical systems: the
one dimensional simple harmonic oscillator, the Kepler problem, the double pendulum, the magnetic mirror, and the three-body problem. This 
algorithm was used in \cite{PhysRevE.106.065205} to investigate adiabatic invariants of ion dynamics in current sheets. The algorithm was 
further developed in \cite{PhysRevE.106.045307} to discover the conservation laws both numerically and symbolically and in 
\cite{PhysRevE.104.055302} in which force decomposition is considered. An alternative framework has been introduced in 
\cite{2021PhRvR...3d2035H}. Another approach has been introduced in \cite{Wetzel_2020}. Further development was recently 
reported in \cite{liu2023discovering}.

The AI Poincar\'{e} essentially works as follows. First, a numeric solution to the relevant Hamilton equations is obtained. The resulting data 
is subjected to preprocessing to whiten the data, which means transforming it to have zero mean and unit covariance. The dimensions with zero 
covariance are omitted from the data via a principal components analysis (PCA). The Dataset is splitted into two parts, for training and 
validating. Then, a noise of scale $L$ is added to the data. Adding noise is a known technique in data processing, namely in the SGLD 
(Stochastic Gradient Langevin Dynamics) \cite{10.5555/3104482.3104568}; see also \cite{2019arXiv190302334S}. The perturbed data is 
feed-forwarded to a neural network with some hidden layers for training. Finally, the trained data is subjected to PCA again, and the Explained 
Variance Ratios are obtained. These explained ratios are then plotted against the noise. Vanishing or nearly vanishing explained ratios are 
interpreted to represent conserved quantities.  

Although neural networks designed to solve various problems in physics (including dynamical systems) have been around for 
some years, the approach described above is quite young (see e.g., the introductory explanation in \cite{2021PhRvL.126r0604L}) and still under 
development. It combines and uses several ideas and methods already known in statistics and computer science and introduces a promising 
algorithm for detection of conservation laws. It would be interesting to examine it in different frameworks. 

{The rich features of dynamical systems naturally provides ground for deploying AI-based techniques. Following this philosophy and
taking the beauty and success of the AI Poincar\'{e} algorithm into account,} in the present work, we apply the machinery described above to 
investigate bifurcations of particle trajectories. Bifurcation is an important concept in dynamical systems arising in mechanics of particles 
and fluids and many other areas of physics. Roughly speaking, it is a sudden change in the behavior of trajectories as some parameters of the 
system are varied. For definition and review, see \cite{RevModPhys.63.991} {, and in particular, the nice review in 
\cite{boccaletti2018}}. 

Our aim is to provide a new useful AI-based tool for investigating bifurcations. This can be used in addition to the 
traditional tools like bifurcation diagrams, phase portraits, and Poincar\'{e} map to understand the physics of systems 
undergoing such changes. The significance of this work is that it enrich the set of tools for the study of bifurcations in 
dynamical systems with the power of a variety of existing and rapidly developing machine learning techniques. 
{The approach would be particularly helpful in higher dimensional systems. This is because it uses PCA as an essential part 
of the algorithm, and PCA is known for its power in dimensional reduction while preserving the data dispersion 
\cite{Jolliffe_2016, gewers3447755}.}   

To achieve this, we consider two different highly interesting systems, the Duffing equation, and the magnetic reversal with shear components. 
Duffing equation is usually considered as a prototype model of nonlinear dynamical systems, and it has wide range of 
applications in physics and engineering, see e.g., \cite{Wawrzynski_2021, 9780470977859.ch2, doi:10.1142/S0218127417501255} and references 
therein. Similarly, dynamics of charged particles in a magnetic reversal is also of interest in physics (particularly in plasma physics) and 
geophysics; see e.g., \cite{1992JGR....9715011C} and references therein. Both {systems} are still subject of active research. 

We show that there are changes in the behavior of the explained variance ratios at bifurcation locations and this can be thought of as 
signaling bifurcations. This is the main contribution of this work. Deep learning methods have recently been applied to study 
conservations laws from the equations of motion as mentioned above, but to the best of my knowledge, they have not been applied to study 
bifurcations. This is the novelty of the present work.

In the following sections, we first describe our method. Then, we give a brief review of the Duffing equation, and for the unperturbed Duffing 
equation we obtain the explained variance ratios for various values of the relevant parameter. We then consider the magnetic reversal with shear 
components and do a similar analysis for some values of the parameters. {We then summarize the results and conclude by a 
discussion of the approach and the results.}    

\section{The method}
We have a set of equations of motion of the form
\begin{equation}\label{dif1}
{\dot{\mathbf X} }=f({\mathbf X}, t)
\end{equation}
for a phase space vector ${\mathbf X}=(X(t), {\dot X}(t))$, some function $f$, and certain initial conditions. Here, 
${\dot X}\equiv\frac{dX}{dt}$. We integrate these equations using a fourth-order Runge-Kutta method with step value of $0.01$. The resulting 
set has $2\times 10^4$ points. It is divided into two parts of equal size (odd and even-labeled points, respectively) to be used for training 
and validation. After preprocessing (i.e., whitening and PCA as described earlier), the training dataset is feed-forwarded to a neural network.
The network has three hidden layers each with 256 nodes, with Leaky ReLU activation functions, mean square loss function MSE, 
Adam optimizer, batch size of $128$ (chosen among a few other options after some experiments), learning rate of $10^{-3}$, number of epochs of 
$2000$, and number of walks of $2000$. {Finally, PCA is applied again, and the resulting} explained variance ratios are 
depicted for noise scales $L=0.1\times j$ for $j=1,2,\cdots, 10$. The neural network is implemented using PyTorch.   

\section{The Duffing equation}\label{duffing}
\subsection{The equation}
The Duffing equation describes a damped anharmonic one-dimensional oscillator given by 
\begin{equation}\label{duf1}
{\ddot x}+\delta{\dot x}+x(x^2-1)(x^2-a)-bx\cos(\Omega t) = \gamma\cos(\omega t)
\end{equation}  
in which the constants $\delta, b, \Omega, \gamma,$ and $\omega$ represent the effects of damping, strength and frequency of the parametric 
excitation, and strength and frequency of the external force, respectively, and $a$ is a free parameter. This equation has been studied 
extensively both for its mathematical properties and its applications in engineering. It has also been studied from a quantum 
mechanical point of view, e.g., in \cite{2021PhRvE.104b4206M}.
By setting $\delta=b=\gamma=0$, this reduces to the unperturbed Duffing equation. We have
\begin{equation}\label{duf2}
{\ddot x}+x(x^2-1)(x^2-a) = 0
\end{equation}
which can be integrated to give
\begin{equation}\label{ee1}
{\dot x}^2=-\frac{1}{3}x^6+\frac{1+a}{2}x^4-ax^2+C
\end{equation}
with $C$ being a constant.

\subsection{The case $a=0$}
We first consider the case $a=0$ with the first integral
\begin{equation}\label{duf2a}
{\dot x}^2=\frac{1}{6}\,x^4\,(3-2\,x^2) + C
\end{equation} 
in which $C\geq -\frac{1}{6}$. By choosing the initial value $(x(0),{\dot x}(0))=(l,k)$, we obtain the constraint
\begin{equation}\label{duf2b}
k = \pm\sqrt{C+\frac{1}{6}\,l^4\,(3-2\,l^2)}.
\end{equation}

The behavior of the solution is sensitive to the value of $C$ and the initial values. For $C=0$, the solution with $(l, k)=(0, 0)$ is the single
point $(x(t),{\dot x}(t))=(0,0)$. The point $(0, 0)$ is called a fixed point, or more specifically, a degenerate saddle. There are two other 
fixed points, $(\pm 1, 0)$ obtained from $C=-\frac{1}{6}$. 

For $C=0$, we have ${\dot x}=0$ at $x=\pm\, {\sqrt\frac{3}{2}}\,\approx\, \pm\, 1.2247$. Thus, a particle initially sitting at these points 
moves towards the origin where it stops. For a particle initially close to the origin in the positive (negative) side
with a positive (negative) velocity, the particle reaches the point $x={\sqrt\frac{3}{2}}$ ($x=-{\sqrt\frac{3}{2}}$) and then continues towards
the origin as shown in Fig. \ref{cap1}.

For $C>0$, orbits are connected as shown in Fig. \ref{cap7}. Such orbits enclose all three fixed points at $x=0,\pm\, 1$. By changing 
$k\leftrightarrow -k$, the same trajectory is obtained.

For $-\frac{1}{6}<C<0$, there are two trajectories in Fig. \ref{fd1} {(plot \ref{cap2})}, each enclosing one of the fixed 
points at $x=\pm\, 1$. Again, changing the sign of $k$ leads to the same plots. These different trajectories are also depicted in 
Fig. \ref{co1} in a single plot for comparison. 

The corresponding explained variance ratios are plotted in Fig. \ref{fd1}. For convenience, related plots are grouped by the color of their
titles. They show a change in the behavior when $C=0$ is crossed. In fact, this corresponds to a bifurcation at $(x(0), {\dot x}(0))=(0,0)$. 
For $C<0$ (plots \ref{cap4} and \ref{cap6})), the difference between component ratios increases up to middle scale noises $L\sim 0.5$, and 
decreases afterwards. For $C=0$ (plots \ref{cap3} and \ref{cap5}), the difference increases up to low to middle scale noises $L\sim 0.3-0.5$, 
and decreases afterwards. For $C>0$ (plot \ref{cap8}), it increases up to middle noise scales $L\sim 0.5$ and remains almost constant
afterwards. 

\subsection{The case $a=-1$}
Now, we consider the case $a=-1$. It does not introduce additional fixed points, but is expected to alter the behavior at $(0,0)$. We obtain 
\begin{equation}\label{ed1}
{\dot x}^2=-\frac{1}{3}\,x^6+x^2+C.
\end{equation}   
Again, the point $(x(t), {\dot x}(t))=(0,0)$ is a solution for $C=0$, and the points $(x(t), {\dot x}(t))=(\pm 1, 0)$ are solutions for 
$C=-\frac{2}{3}$. The phase portraits and the explained variance ratios are shown in Fig. \ref{fd3} for some relevant values of $C$ and 
the initial conditions. All trajectory types are shown in Fig. \ref{co3} in a single plot. 

The corresponding explained ratio plots show a change of behavior when $C$ is varied. For $C<0$ (plots \ref{cap14}, \ref{cap16}), the ratios 
difference increases up to $L\sim 0.2-0.5$ and decreases afterwards. For $C>0$ (plot \ref{cap18}), it increases up to $L\sim 0.2$ and then 
decreases. For $C=0$ (plots \ref{cap13} and \ref{cap15}), it increases up to $L\sim 0.2$ and remains almost constant afterwards.     

\subsection{The case $a=1$}
In this case, we have
\begin{equation}\label{ed2}
{\dot x}^2=-\frac{1}{3}\,x^6+x^4-x^2+C.
\end{equation} 
For $C=0$, the point $(x(t),{\dot x}(t))=(0,0)$ is a solution. For $C=\frac{1}{3}$, the points $(x(t),{\dot x}(t))=(\pm 1, 0)$ are
solutions. For $0<C<\frac{1}{3}$, there is a trajectory (plot \ref{cap21}) enclosing the point $(0,0)$, but not the points $(\pm 1, 0)$. For 
$C=\frac{1}{3}$, there is a path emanating from a neighborhood of $(1,0)$ ending at $(-1,0)$ and a similar path in the reverse direction 
(plot \ref{cap25}). For $C>\frac{1}{3}$, there is trajectory (plot \ref{cap22}) enclosing all the three fixed points. These trajectories and 
the corresponding explained ratios are shown in Fig. \ref{fd5} for typical values of the constant and starting points. 
All of these trajectories are plotted together in Fig. \ref{co2}. 
\begin{figure}
  \centering
  \captionsetup[subfigure]{skip=-55mm}
  \subcaptionbox{\raggedright\label{cap1}}{\includegraphics[scale=.5]{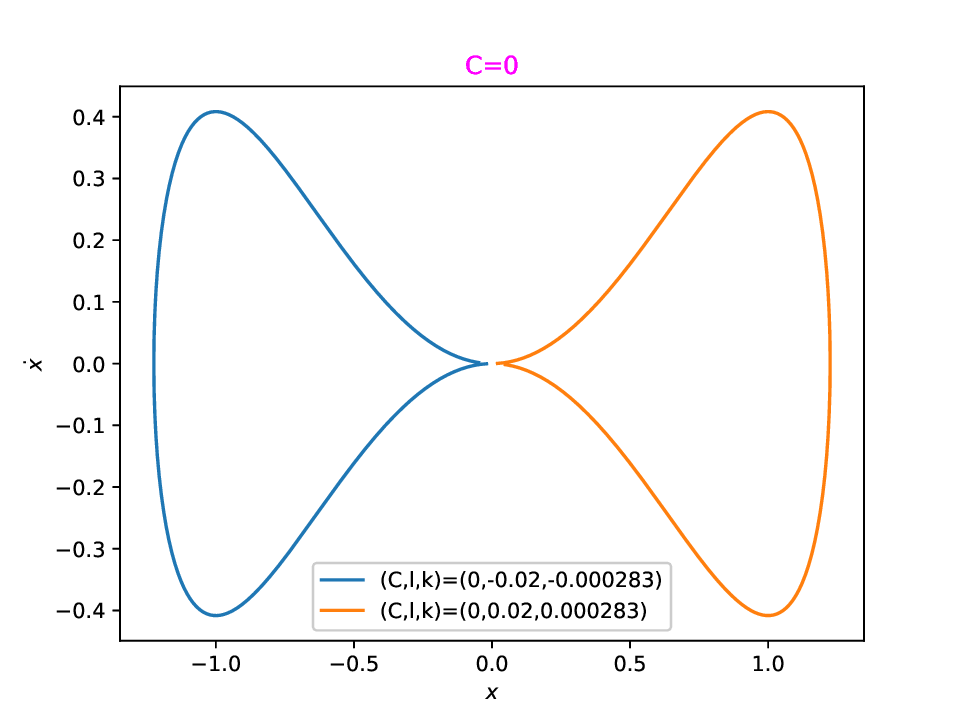}}
  \subcaptionbox{\raggedright\label{cap2}}{\includegraphics[scale=.5]{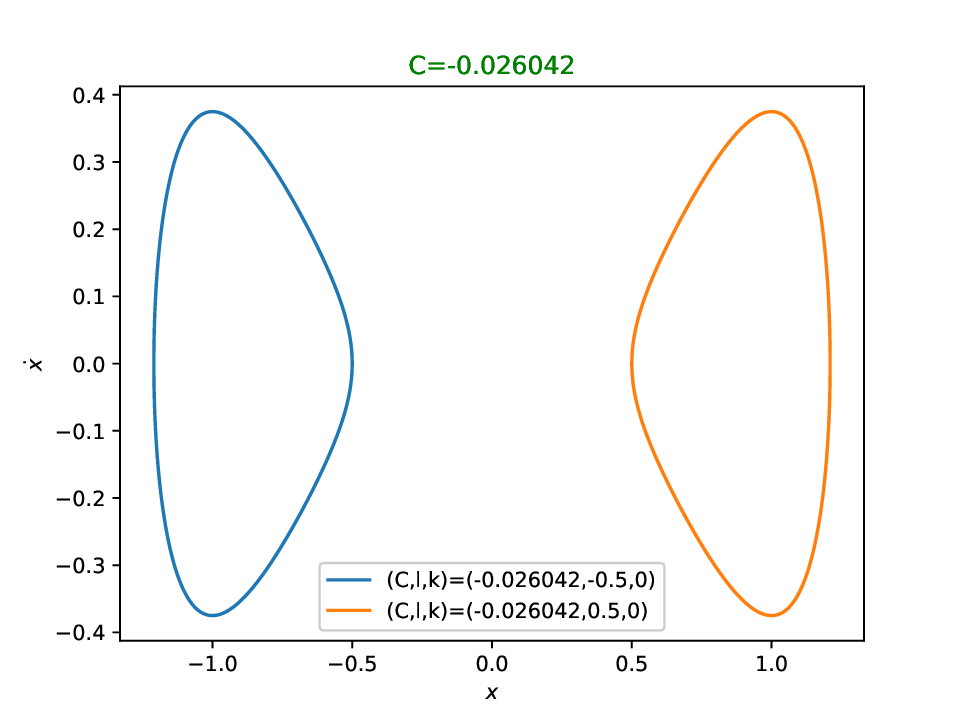}}
  \subcaptionbox{\raggedright\label{cap3}}{\includegraphics[scale=.5]{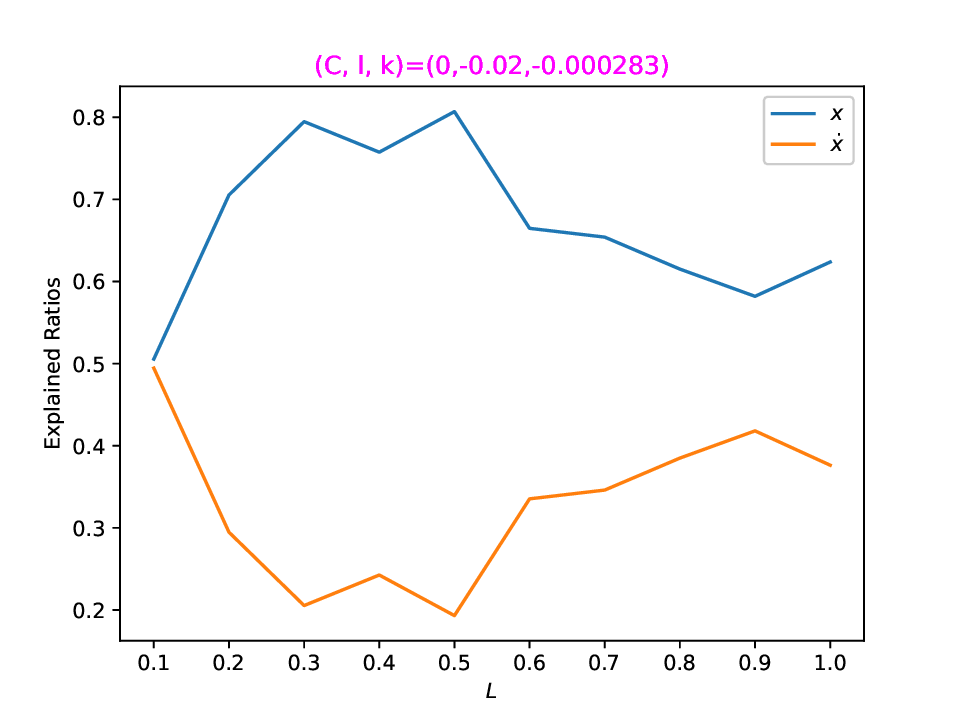}}
  \subcaptionbox{\raggedright\label{cap4}}{\includegraphics[scale=.5]{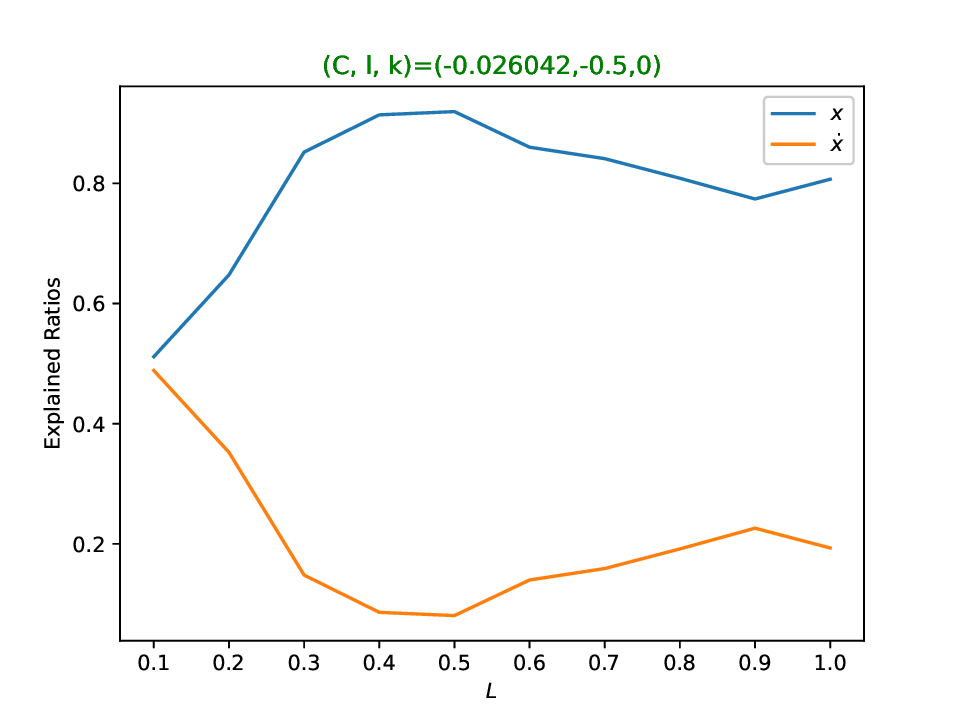}}
  \subcaptionbox{\raggedright\label{cap5}}{\includegraphics[scale=.5]{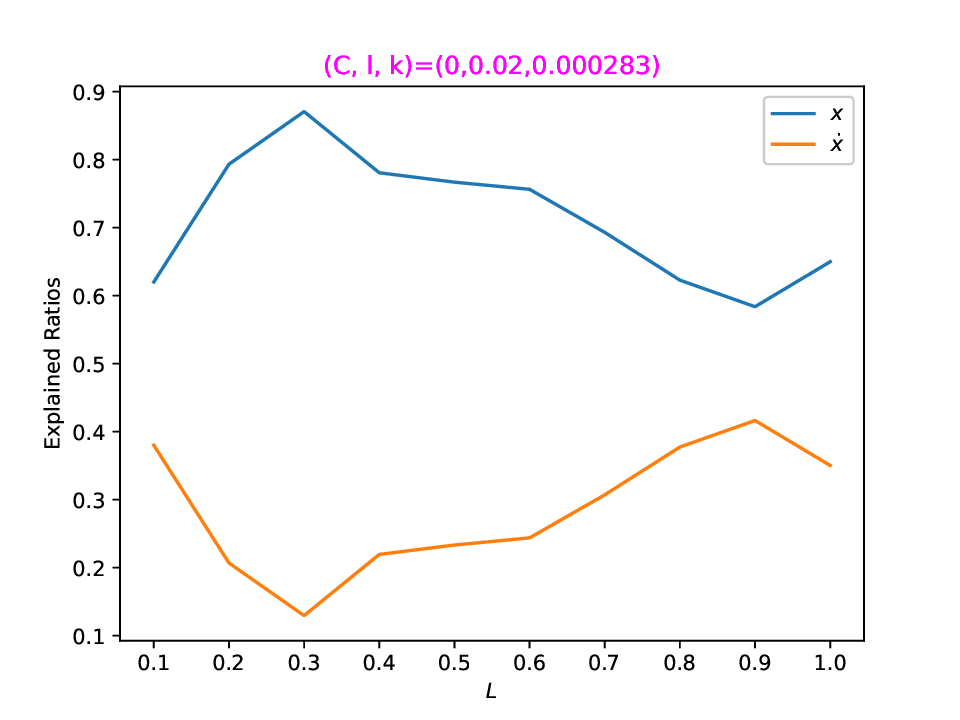}}
  \subcaptionbox{\raggedright\label{cap6}}{\includegraphics[scale=.5]{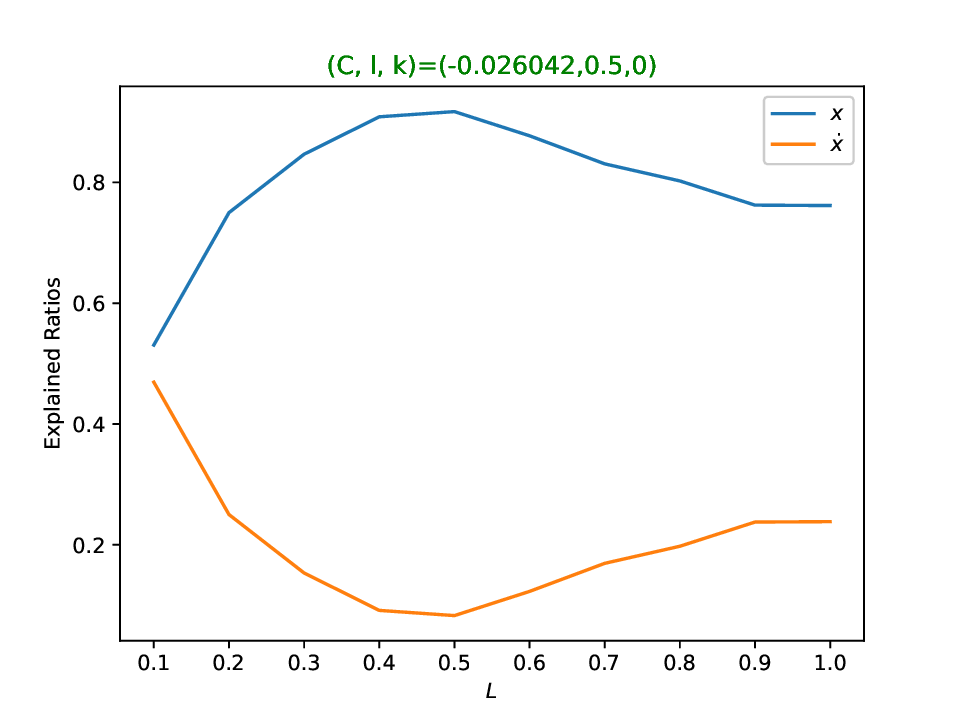}}
  \caption{The phase portraits and explained variance ratios for the unperturbed Duffing equation with $a=0$ for $C=0$, $l=\pm 0.02$, 
  $k=\pm\sqrt{\frac{0.02^4}{2}-\frac{0.02^6}{3}}\, \approx\,\pm\,0.000283$ (plots \subref{cap1}, \subref{cap3}, and \subref{cap5}), 
  $C= -\left(\frac{0.5^4}{2}-\frac{0.5^6}{3}\right)\, \approx\,-0.026042$, $l= \pm 0.5$, $k=0$ (plots \subref{cap2},
  \subref{cap4}, and \subref{cap6}), and $C=0.0025$, $l=0$, $k=0.05$ (plots \subref{cap7} and \subref{cap8}) .} \label{fd1} 
  \end{figure}

  Here, the difference between ratios increases up to $L\sim 0.2$ and decreases afterwards for both $C<\frac{1}{3}$ (plot \ref{cap23}) and 
$C>\frac{1}{3}$ (plot \ref{cap24}). For $C=\frac{1}{3}$ (plots \ref{cap26} and \ref{cap27}) it first increase up to $L\sim 0.2$ and then 
fluctuates around an average value. A change of behavior is seen when $C=\frac{1}{3}$ is crossed.

\subsection{The case $0<a<1$}
For $0<a<1$, the situation is more complicated as this introduces two more fixed points. Let us choose a typical value, say, $a=\frac{1}{3}$. 
We obtain
\begin{equation}
{\dot x}^2=-\frac{1}{3}\,x^6+\frac{2}{3}\,x^4-\frac{1}{3}\,x^2+C.    
\end{equation}
Choosing $C=0$, the points $(0,0), (\pm 1, 0)$ are solutions. For $C=\frac{4}{81}$, there are two other points $(\pm\sqrt{\frac{1}{3}},0)$ as 
solutions. Thus, there are five fixed points altogether. Depending on the values of $C$, different trajectories are possible as shown in 
Fig. \ref{fd7}. For $0<C<\frac{4}{81}$, there are trajectories enclosing the points $(0,0)$ (plot \ref{cap31}), $(-1,0)$, and $(1,0)$ 
(plot \ref{cap35}), respectively. For $C=\frac{4}{81}$, there are two trajectories emanating from  $(-\sqrt{\frac{1}{3}}+\epsilon,0)$
($\epsilon$ being small and positive) and ending at $(\sqrt{\frac{1}{3}},0)$ (plot \ref{cap32}) and similar trajectories in 
the reverse direction. There is also a trajectory emanating from a neighborhood of these points and ending at themselves (plot \ref{cap36}). 
Finally, for $C>\frac{4}{81}$ there is a trajectory enclosing all five fixed points (plot \ref{cap39b}). 
These trajectories are depicted in a single plot in Fig. \ref{co4}.

The corresponding explained ratios are depicted in Fig. \ref{fd7}. For $C<\frac{4}{81}$ and the starting points close to $x=0$ or $x=\pm 1$
(plots \ref{cap33}, \ref{cap37}, \ref{cap39}), the ratios difference is ascending to $L\sim 0.4-0.5$ and descending afterwards. For 
$C=\frac{4}{81}$ and the starting point in the interval $-\frac{1}{\sqrt 3}<x<\frac{1}{\sqrt 3}$ (plot \ref{cap34}), the 
difference behavior has the pattern ascending-descending-ascending. For $C=\frac{4}{81}$ and the starting point in $|x|>\frac{1}{\sqrt 3}$ 
(plots \ref{cap38}, \ref{cap39a}), the difference increases up to $L\sim 0.2$ and remains almost constant afterwards. Finally, for 
$C>\frac{4}{81}$ (plot \ref{cap39c}), the difference in ratios is ascending up to $L\sim 0.2$ and descending afterwards. Thus, changes of 
behavior occurs when $C$  crosses $\frac{4}{81}$ and the location of starting point is varied around $x=\frac{1}{\sqrt 3}$.  

  \begin{figure}\ContinuedFloat
  \centering   
  \captionsetup[subfigure]{skip=-55mm}
  \subcaptionbox{\raggedright\label{cap7}}{\includegraphics[scale=.5]{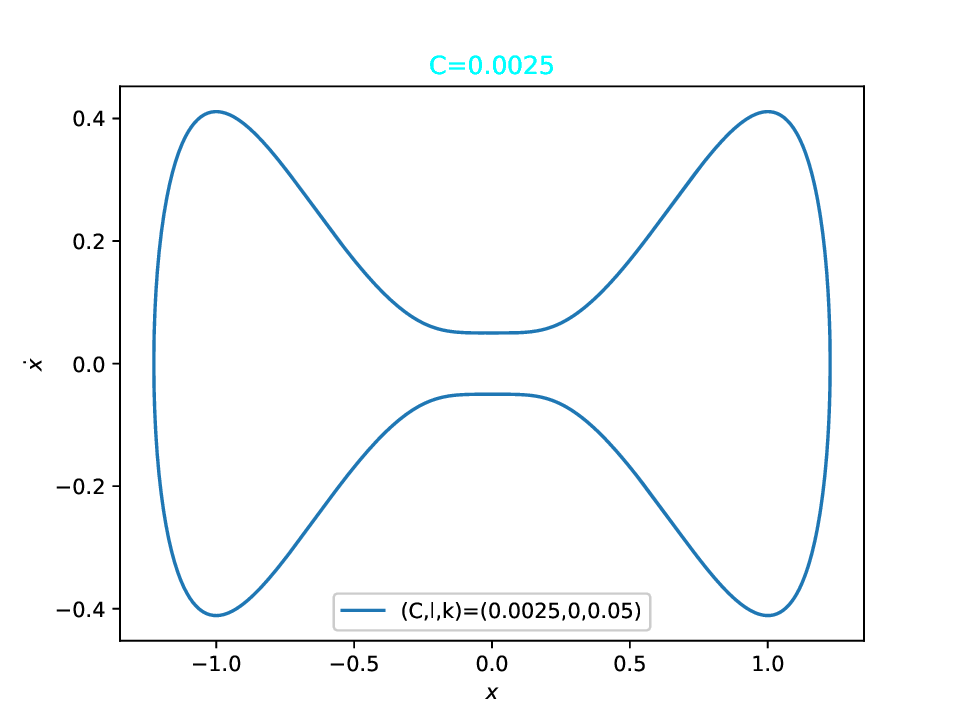}}
  \subcaptionbox{\raggedright\label{cap8}}{\includegraphics[scale=.5]{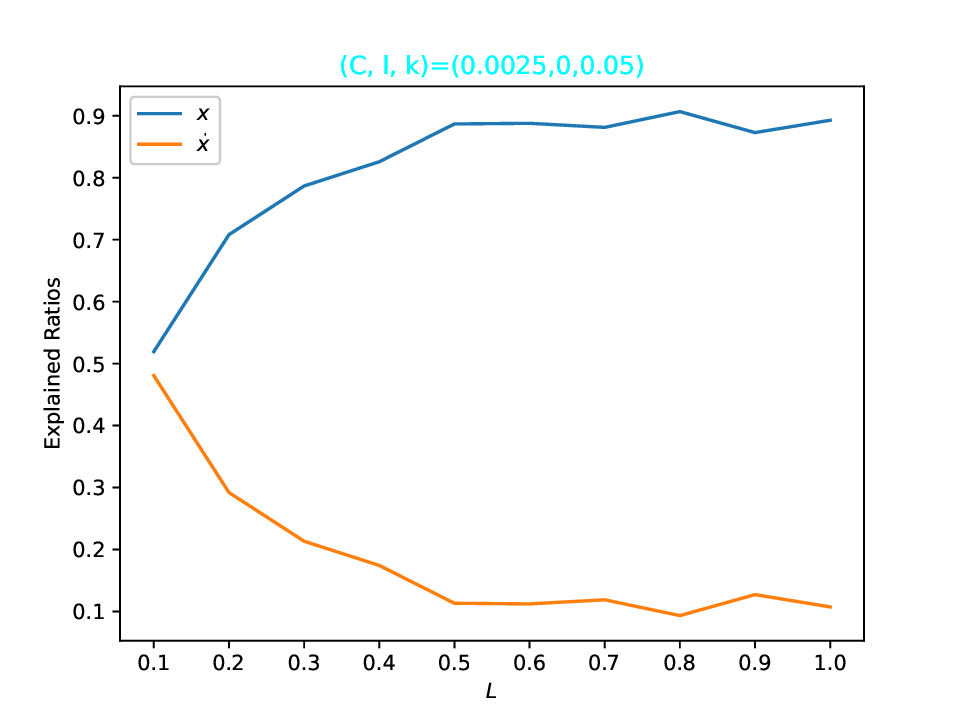}}
  \caption{Continued.}  
  \end{figure}
  \begin{figure}
    \centering
    \captionsetup[subfigure]{skip=-55mm}
    \subcaptionbox{\raggedright\label{cap11}}{\includegraphics[scale=.5]{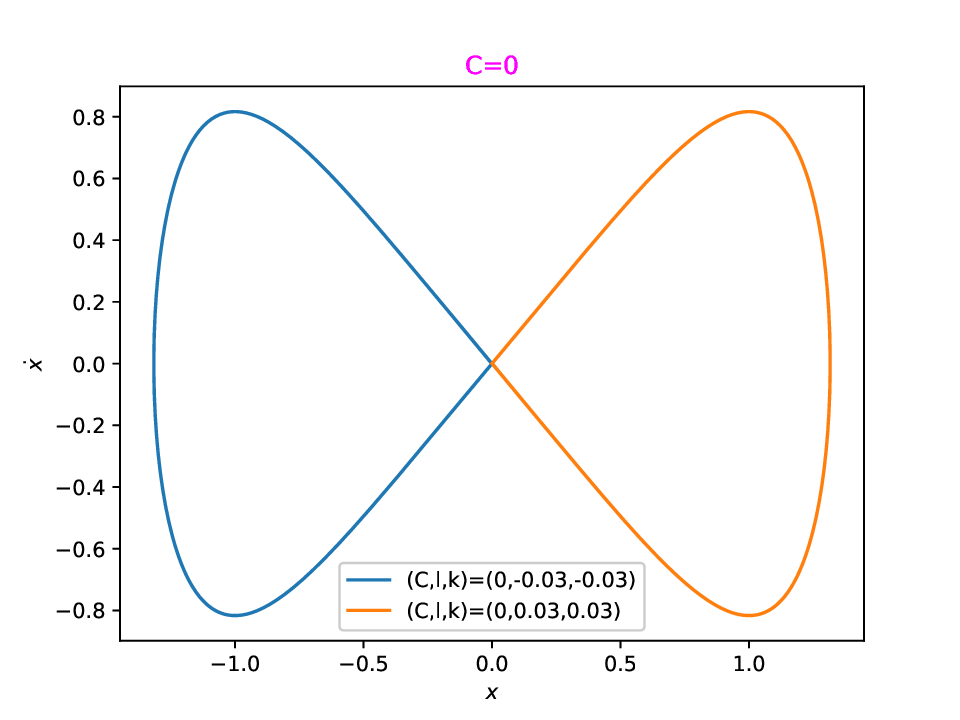}}
    \subcaptionbox{\raggedright\label{cap12}}{\includegraphics[scale=.5]{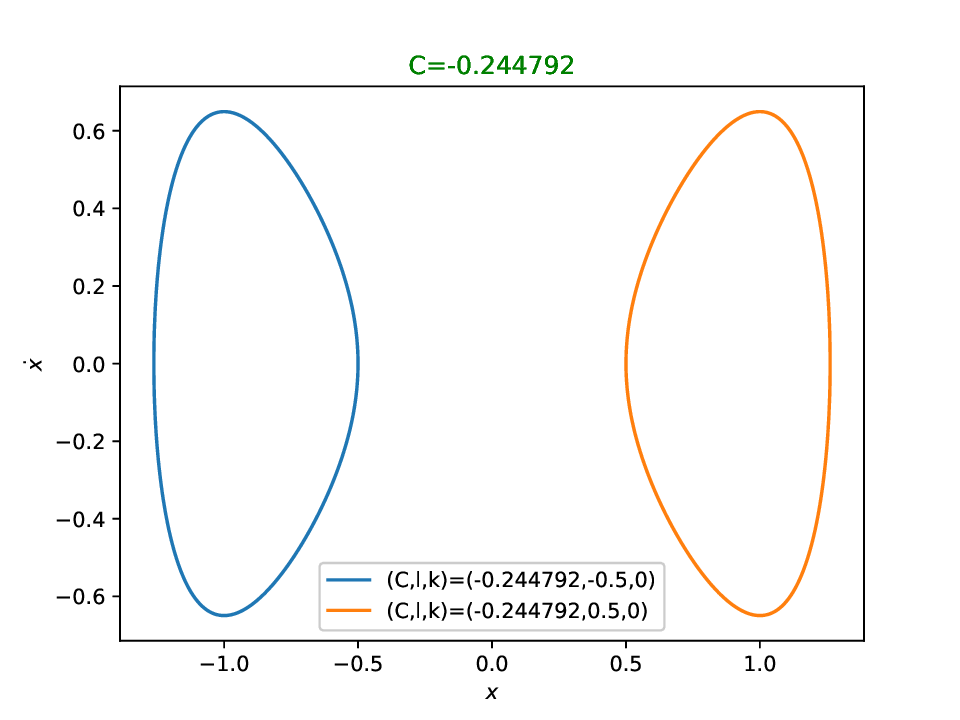}}
    \subcaptionbox{\raggedright\label{cap13}}{\includegraphics[scale=.5]{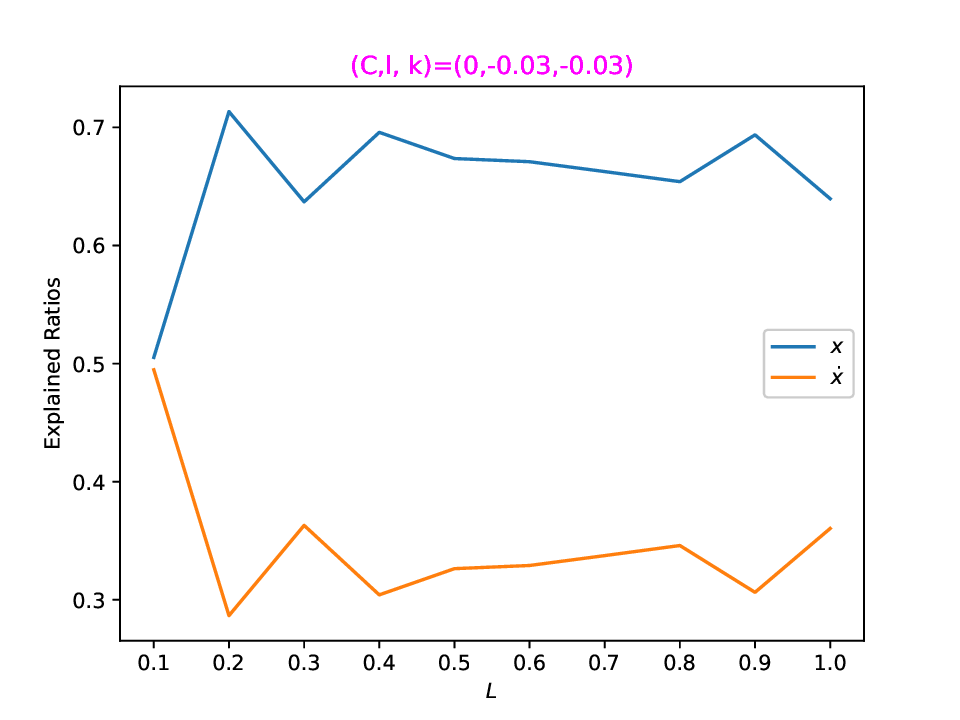}}
    \subcaptionbox{\raggedright\label{cap14}}{\includegraphics[scale=.5]{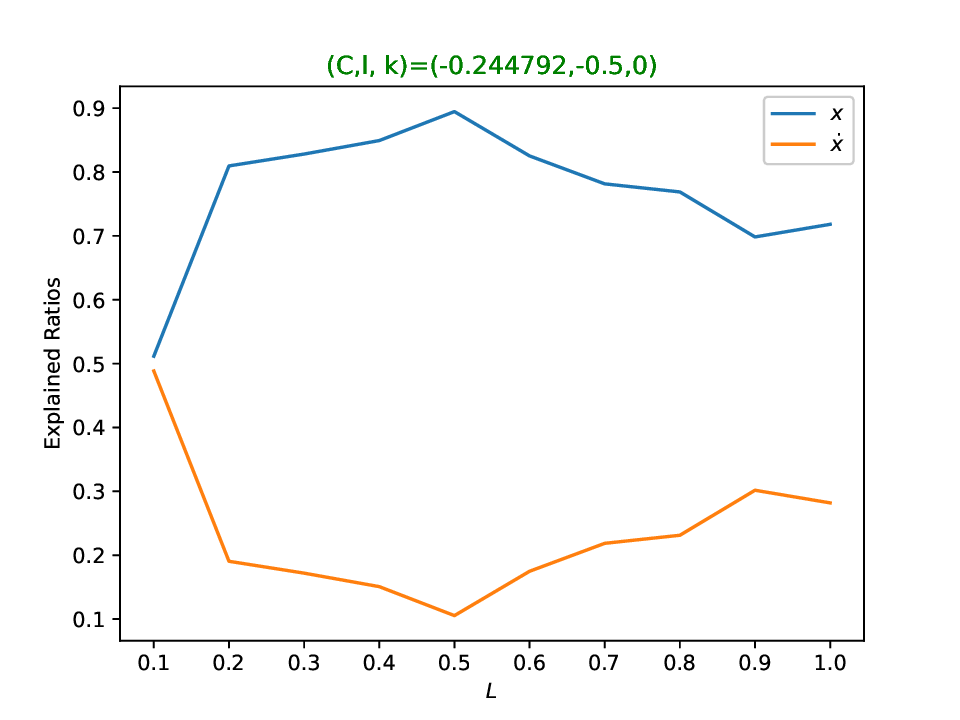}}
    \subcaptionbox{\raggedright\label{cap15}}{\includegraphics[scale=.5]{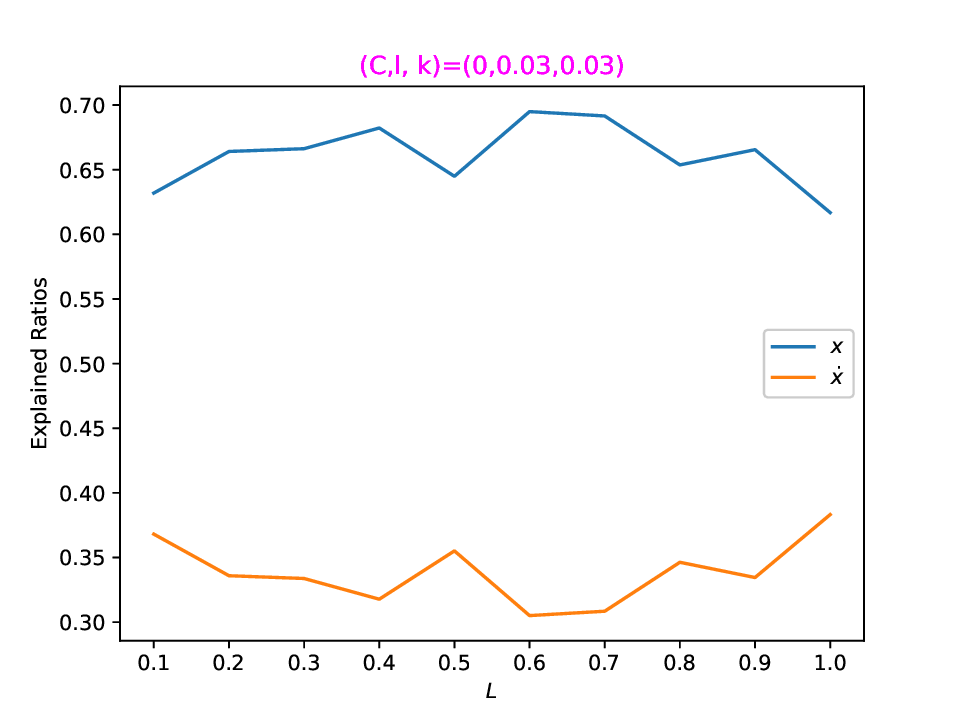}}
    \subcaptionbox{\raggedright\label{cap16}}{\includegraphics[scale=.5]{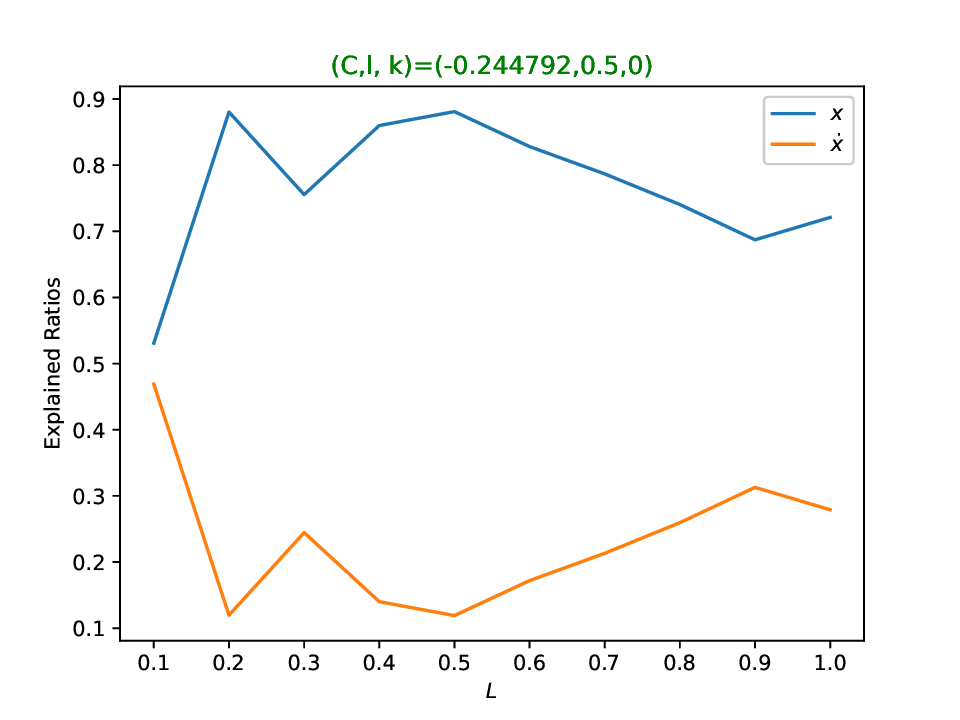}}
    \caption{The phase portraits and explained variance ratios for the unperturbed Duffing equation with $a=-1$ for $C=0$, $l=\pm 0.03$, 
    $k=\pm\sqrt{0.03^2-\frac{0.03^6}{3}}\,\approx\,\pm\, 0.03$ (plots \subref{cap11}, \subref{cap13}, and \subref{cap15}),  
    $C= -\left(0.5^2-\frac{0.5^6}{3}\right)\, \approx\, -0.244792$, $l= \pm 0.5$, 
    $k=0$ (plots \subref{cap12}, \subref{cap14}, and \subref{cap16}), and $C=0.0025$, $l=0$, $k=0.05$ (plots \subref{cap17} 
    and \subref{cap18}). }\label{fd3} 
    \end{figure}
    \begin{figure}\ContinuedFloat
    \centering
    \captionsetup[subfigure]{skip=-55mm}
    \subcaptionbox{\raggedright\label{cap17}}{\includegraphics[scale=.5]{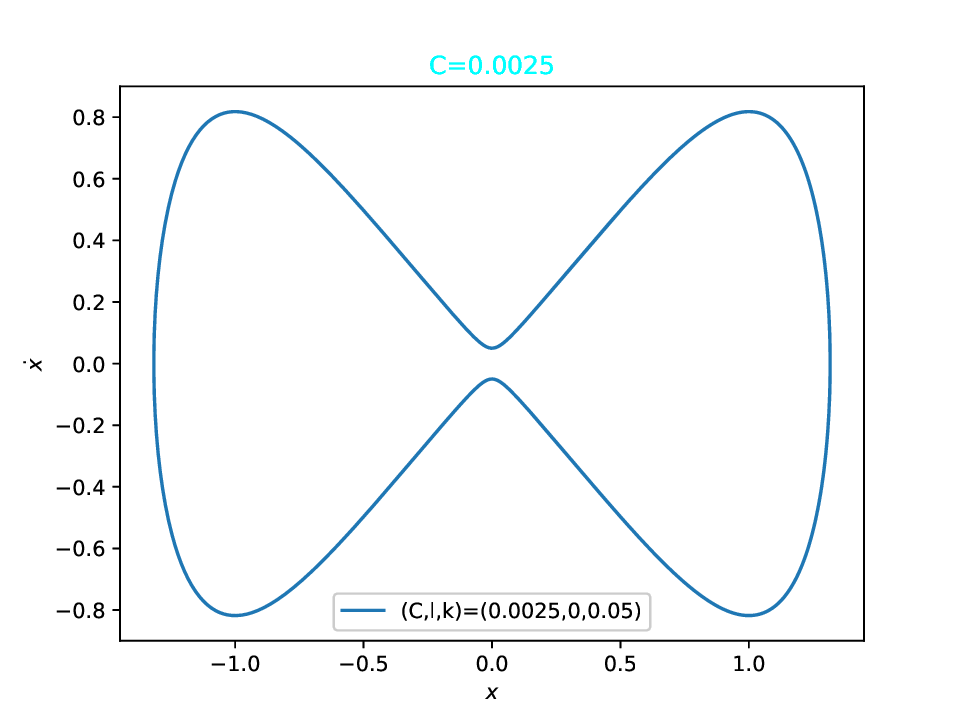}}
    \subcaptionbox{\raggedright\label{cap18}}{\includegraphics[scale=.5]{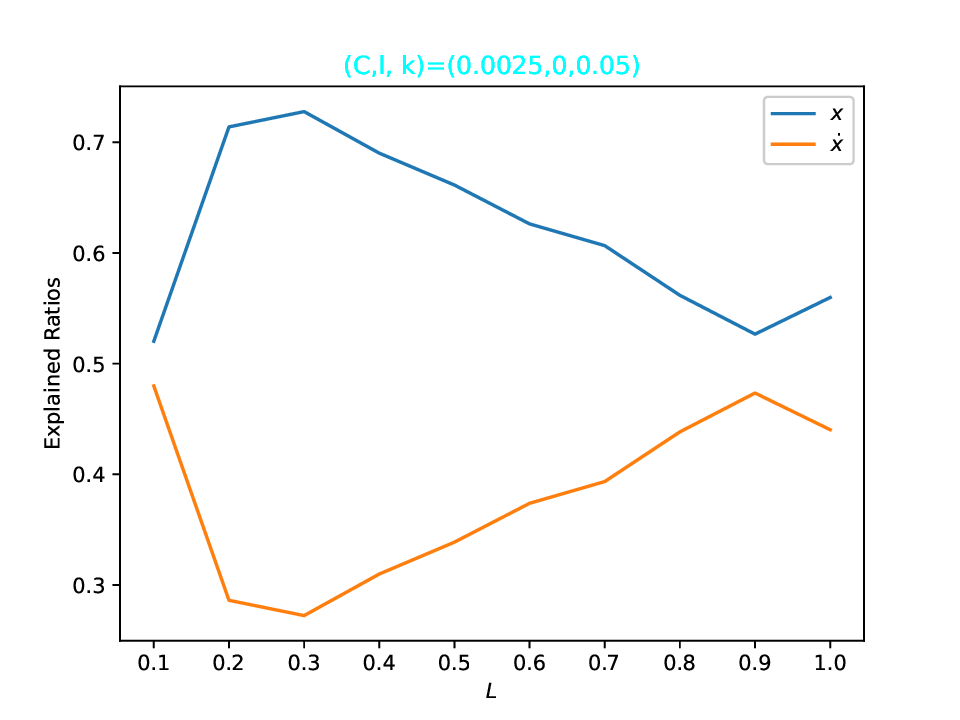}}
    \caption{Continued. } 
    \end{figure} 
    \begin{figure}
      \centering
      \captionsetup[subfigure]{skip=-55mm}
      \subcaptionbox{\raggedright\label{cap21}}{\includegraphics[scale=.5]{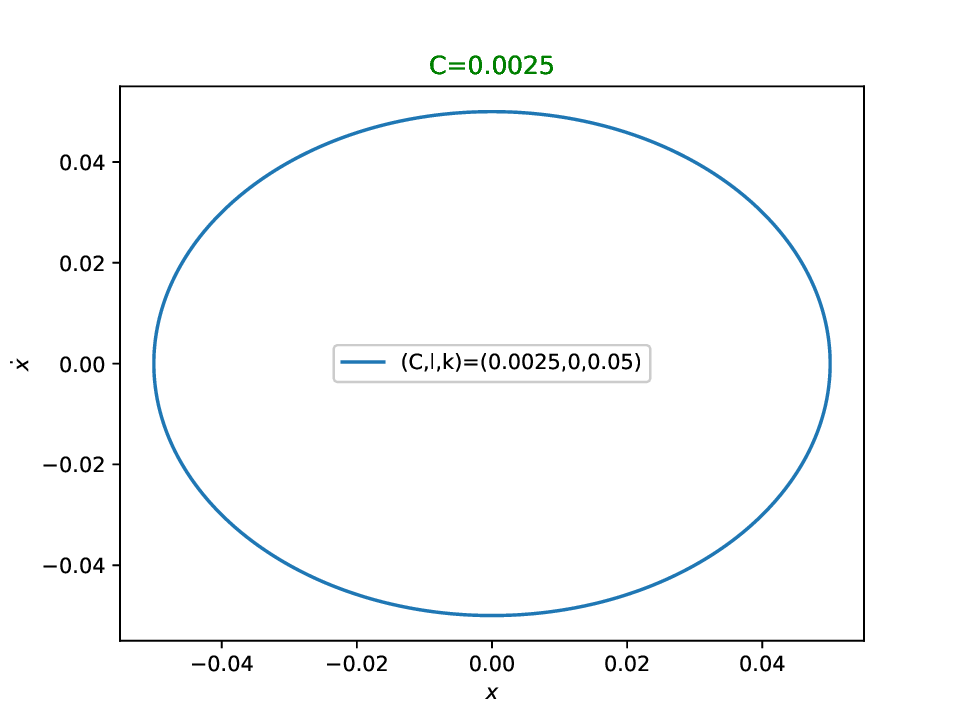}}
      \subcaptionbox{\raggedright\label{cap22}}{\includegraphics[scale=.5]{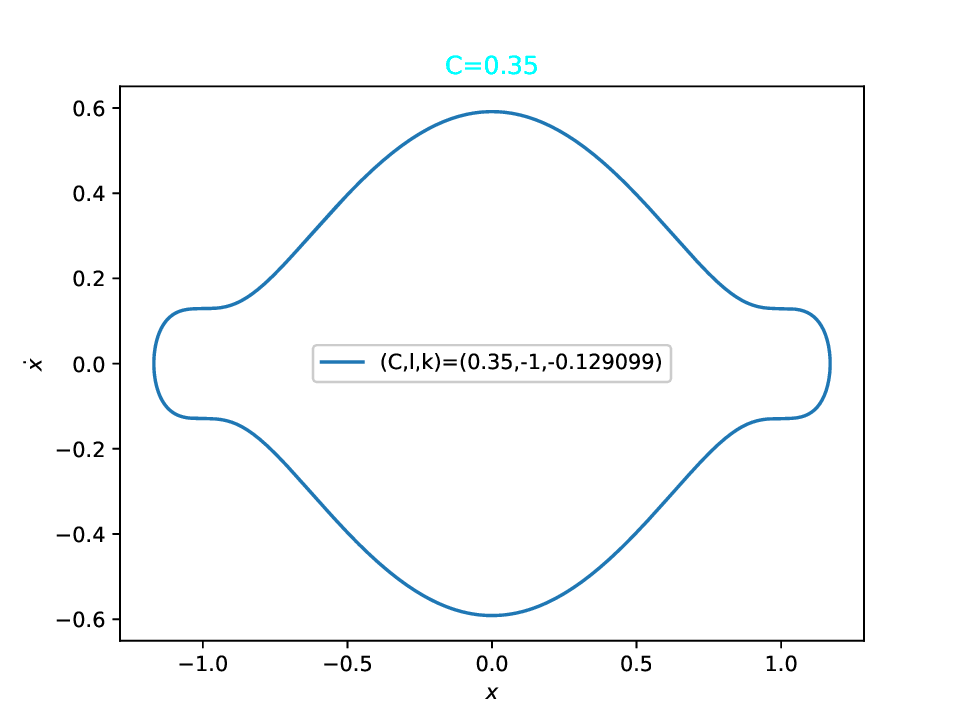}}
      \subcaptionbox{\raggedright\label{cap23}}{\includegraphics[scale=.5]{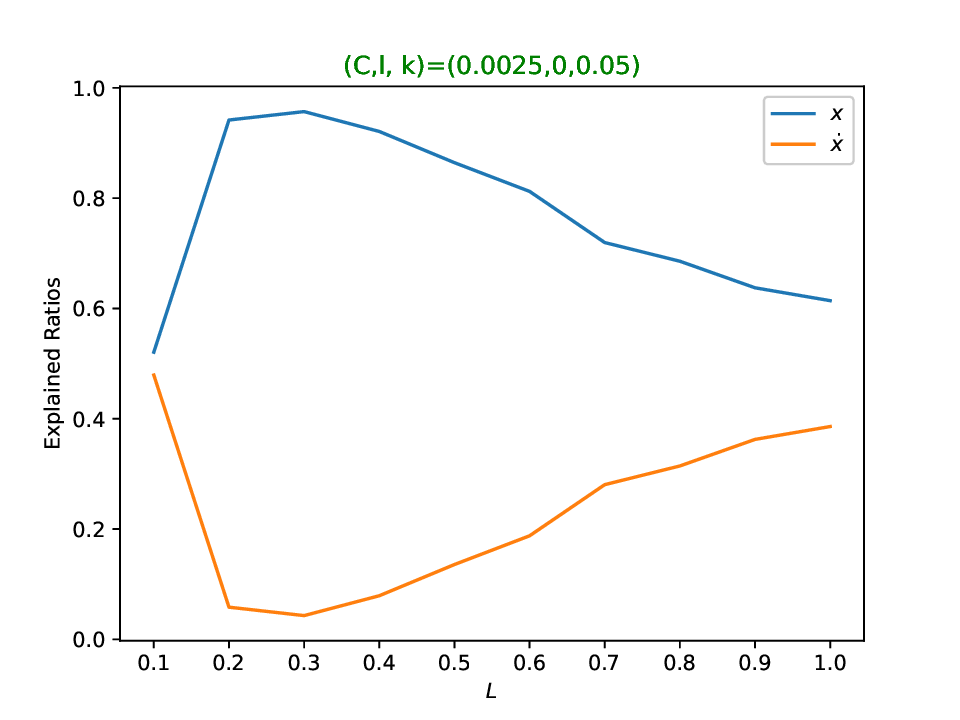}}
      \subcaptionbox{\raggedright\label{cap24}}{\includegraphics[scale=.5]{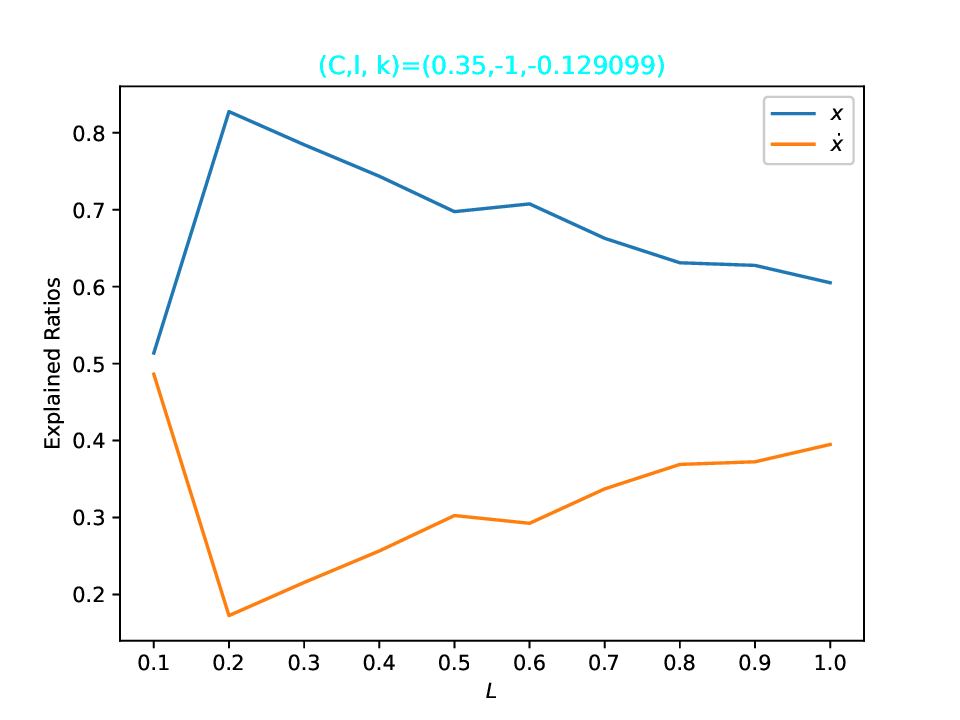}}
      \caption{The phase portraits and explained variance ratios for the unperturbed Duffing equation with $a=1$ for
      $C=0.0025$, $l=0$, $k=0.05$ (plots \subref{cap21} and \subref{cap23}), 
      {$C=0.35$, $l=-1$, $k= -\sqrt{-\frac{1}{3}+0.35}\,\approx\,-0.129099$ 
      (plots \subref{cap22} and \subref{cap24}), and $C=\frac{1}{3}\, \approx\, 0.333333$, $l=\pm 0.98$, 
      $k=\mp \sqrt{-0.98^2+0.98^4-\frac{0.98^6}{3}+\frac{1}{3}}\, 
      \approx\,\mp\, 0.00455$ (plots \subref{cap25}, \subref{cap26}, and \subref{cap27}).}}\label{fd5} 
      \end{figure}
    \begin{figure}\ContinuedFloat
      \centering
      \captionsetup[subfigure]{skip=-55mm}
      \subcaptionbox{\raggedright\label{cap25}}{\includegraphics[scale=.5]{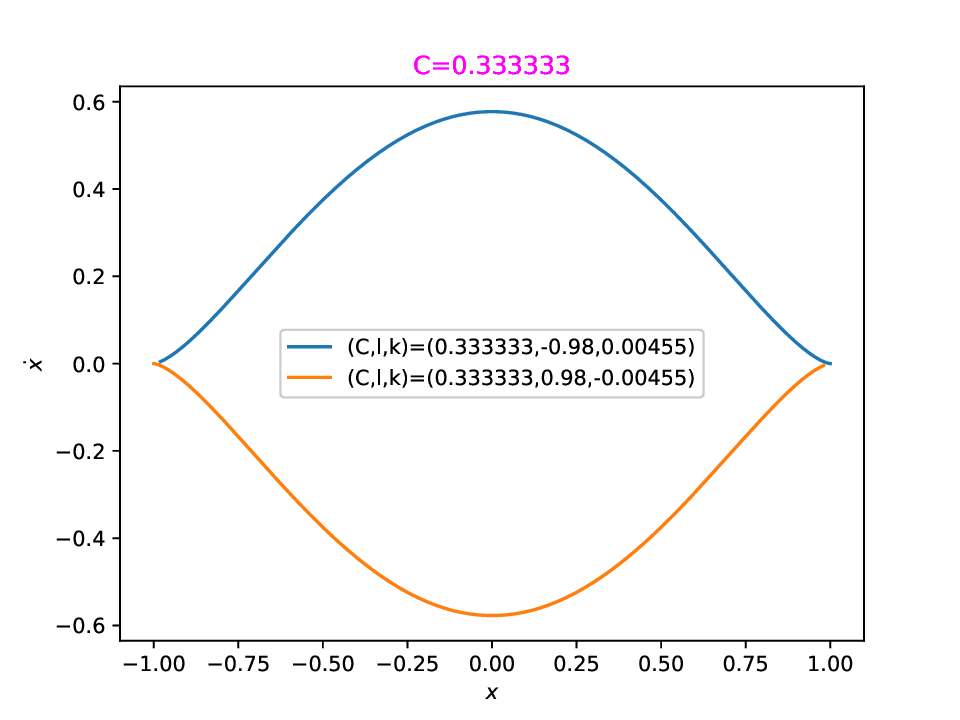}}\\
      \subcaptionbox{\raggedright\label{cap26}}{\includegraphics[scale=.5]{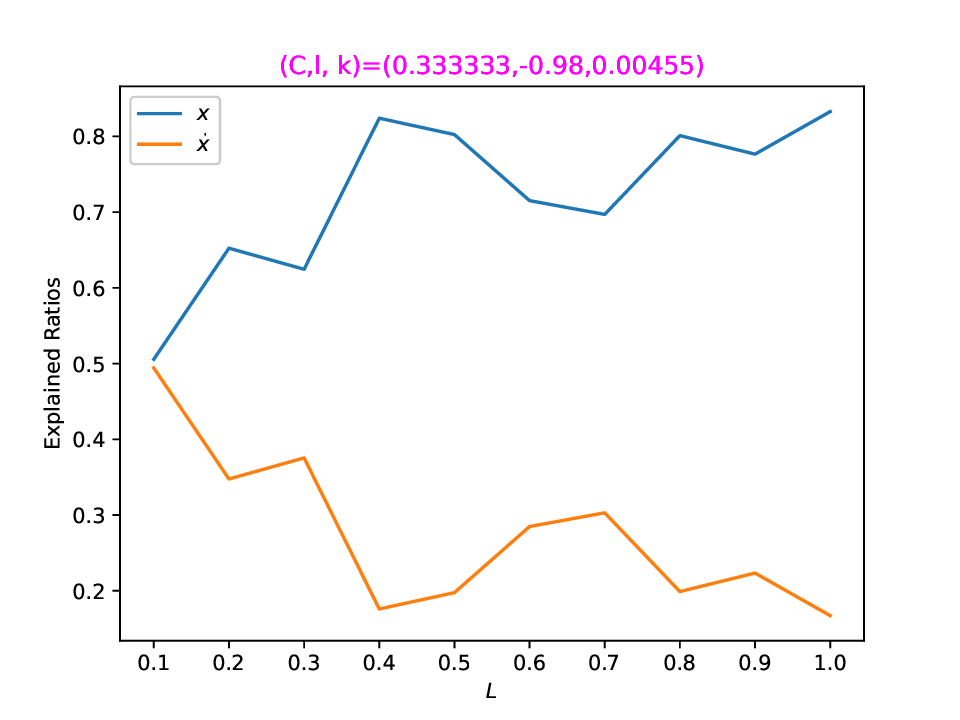}}
      \subcaptionbox{\raggedright\label{cap27}}{\includegraphics[scale=.5]{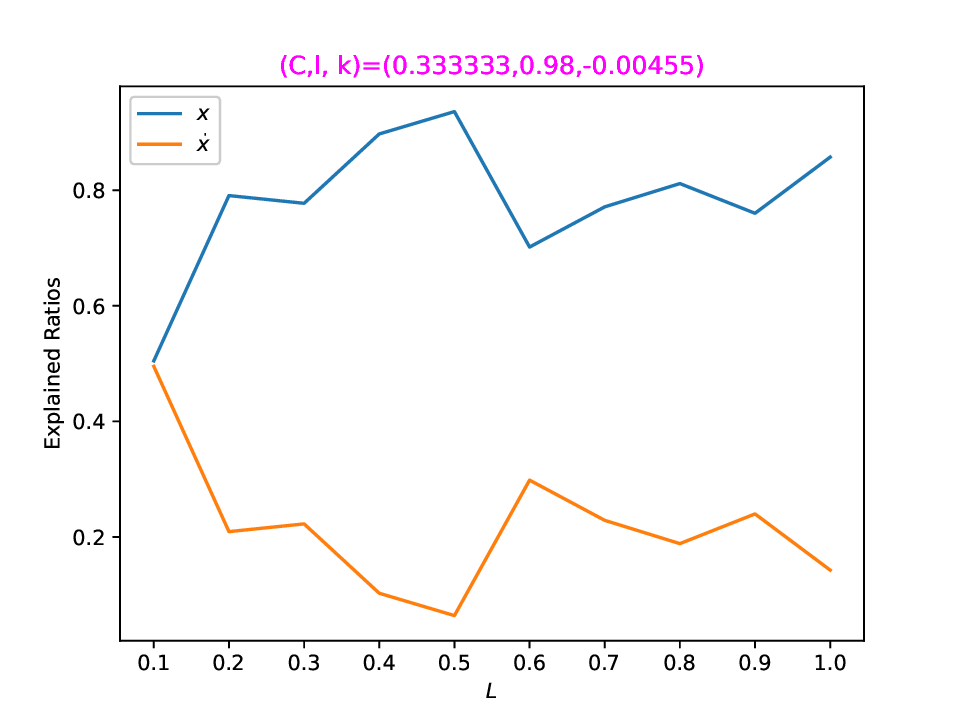}}
      \caption{Continued.} 
      \end{figure}
 
\begin{figure}
\centering
\captionsetup[subfigure]{skip=-55mm}
\subcaptionbox{\raggedright\label{cap31}}{\includegraphics[scale=.5]{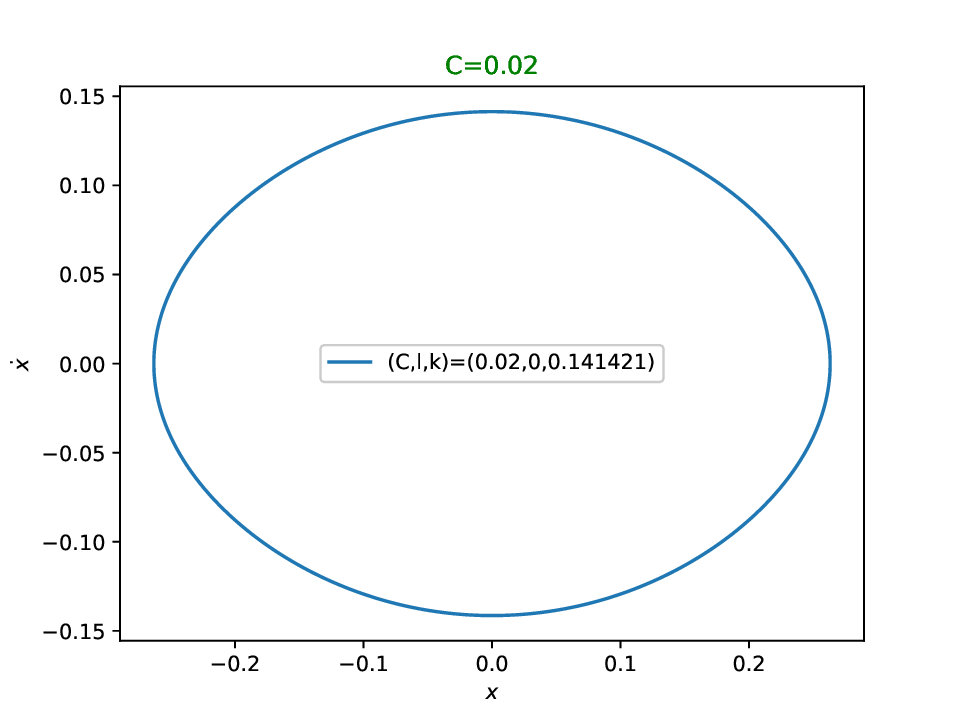}}
\subcaptionbox{\raggedright\label{cap32}}{\includegraphics[scale=.5]{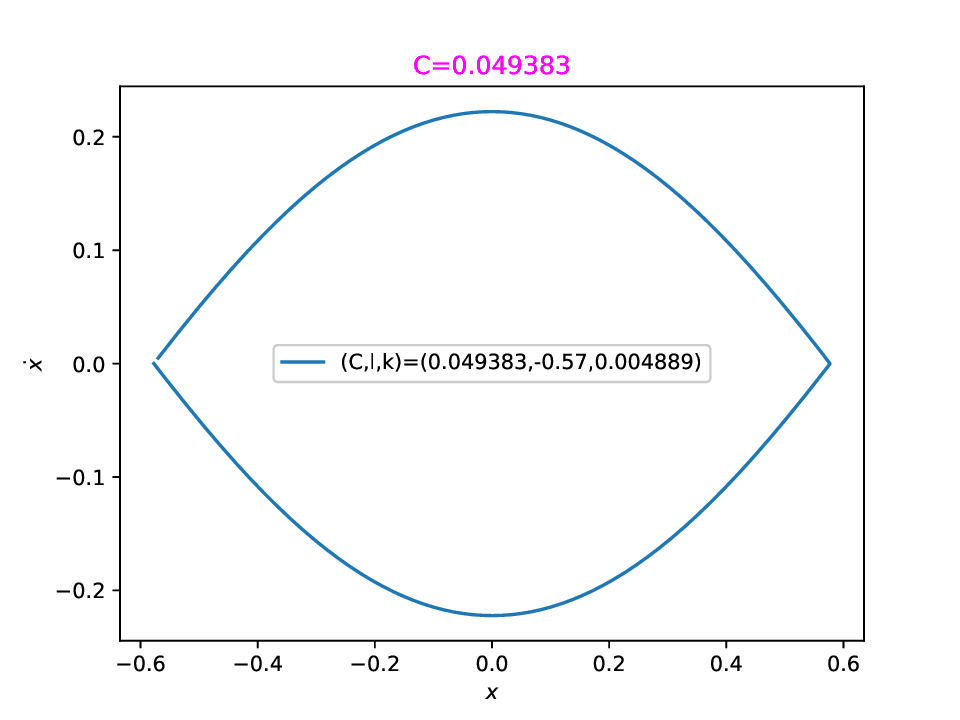}}
\subcaptionbox{\raggedright\label{cap33}}{\includegraphics[scale=.5]{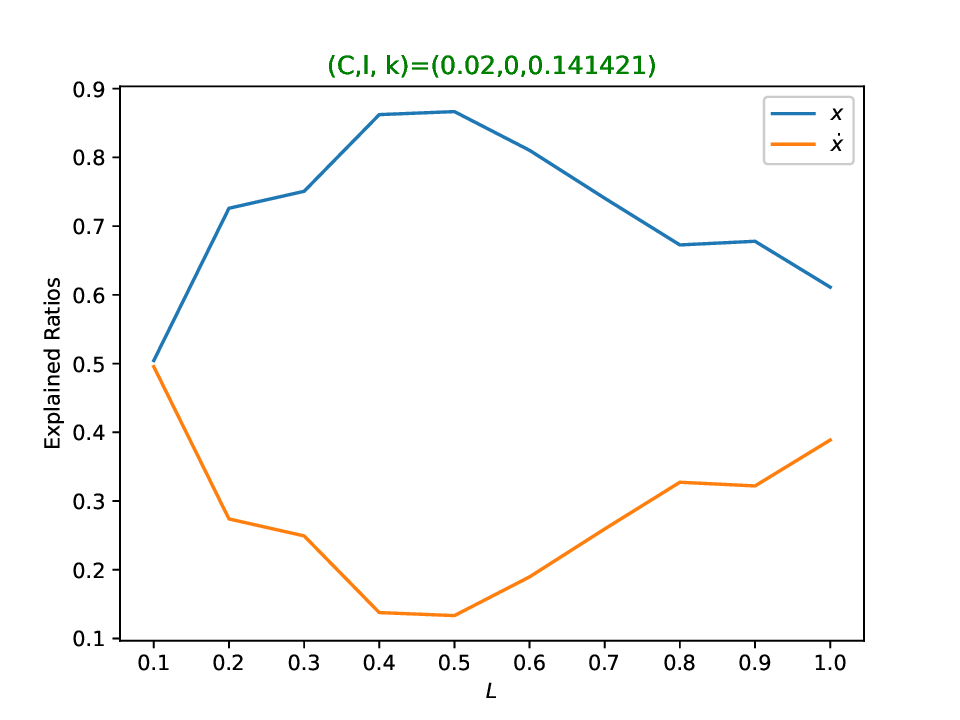}}
\subcaptionbox{\raggedright\label{cap34}}{\includegraphics[scale=.5]{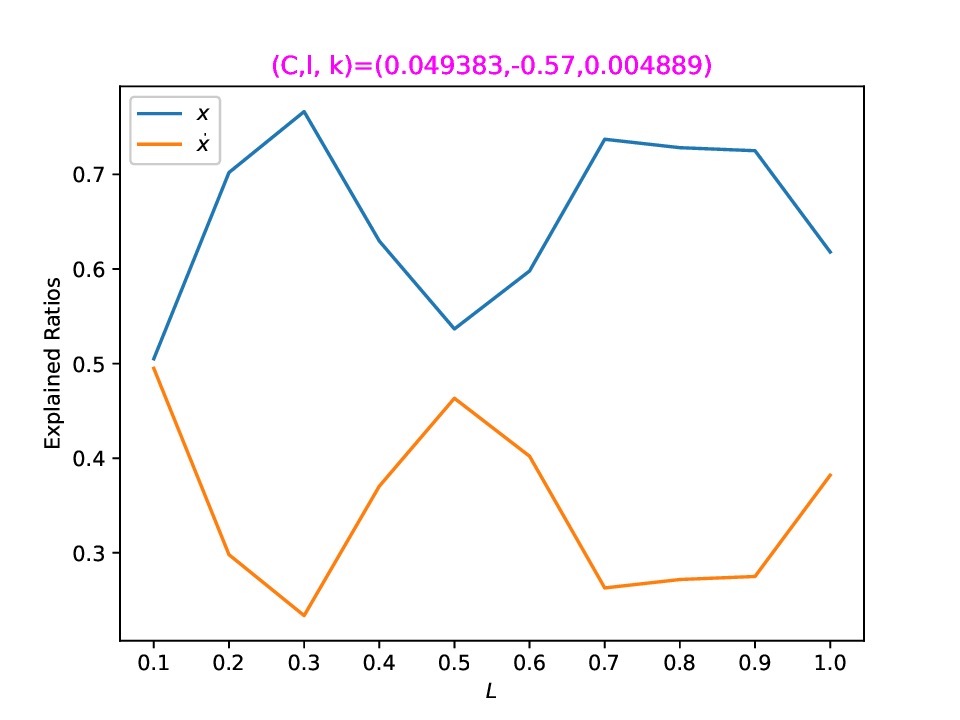}}
\caption{The phase portraits for the unperturbed Duffing equation with $a=\frac{1}{3}$ for 
$C=0.02$, $l=0$, $k=\sqrt{0.02}\approx 0.141421$ (plots \subref{cap31} and \subref{cap33}),  
$C=\frac{4}{81}\approx 0.049383$, $l=-0.57$, 
$k= \sqrt{-\frac{1}{3}(-0.57)^2((-0.57)^2-1)^2+\frac{4}{81}}\approx 0.004889$ 
(plots \subref{cap32} and \subref{cap34}), $C=0.048$, $l=\pm 1$, $k=\mp \sqrt{0.048}\approx \mp 0.219089$ (plots \subref{cap35}, 
\subref{cap37}, and \subref{cap39}), $C=\frac{4}{81}\approx 0.049383$, $l=\pm 0.59$, 
$k= \pm\sqrt{-\frac{1}{3}(0.59)^2((0.59)^2-1)^2+\frac{4}{81}} 
\approx \pm 0.008462$ (plots \subref{cap36}, \subref{cap38}, and \subref{cap39a}), and $C=0.5$, $l=0$, $k= \sqrt{0.5}\approx 
0.707107$ (plots \subref{cap39b} and \subref{cap39c}). }\label{fd7} 
\end{figure}
\begin{figure}\ContinuedFloat
\centering
\captionsetup[subfigure]{skip=-55mm}  
\subcaptionbox{\raggedright\label{cap35}}{\includegraphics[scale=.5]{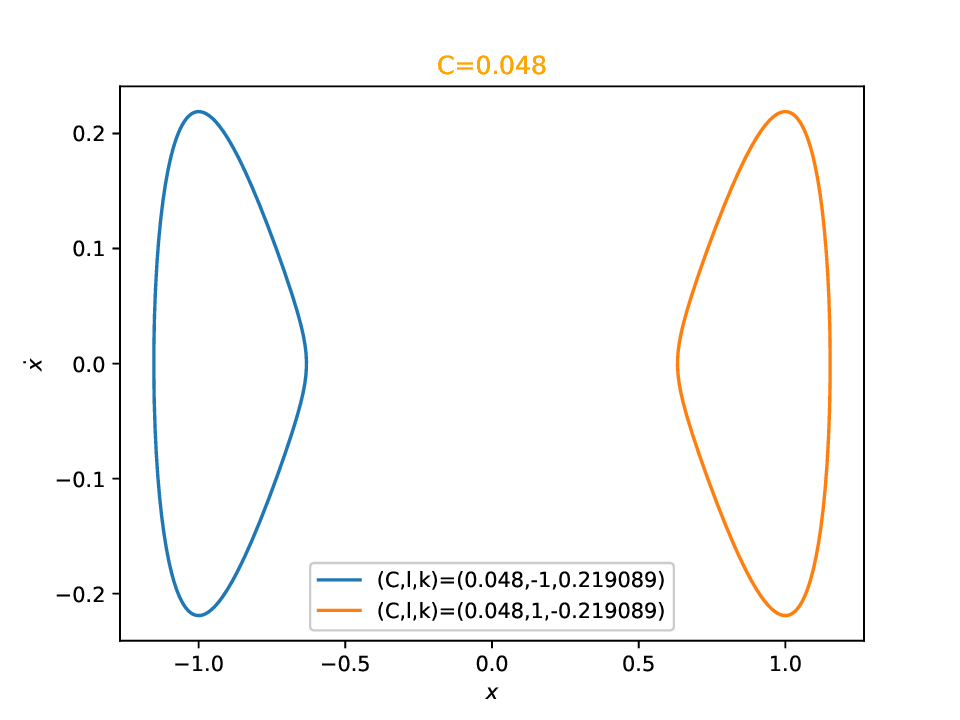}}
\subcaptionbox{\raggedright\label{cap36}}{\includegraphics[scale=.5]{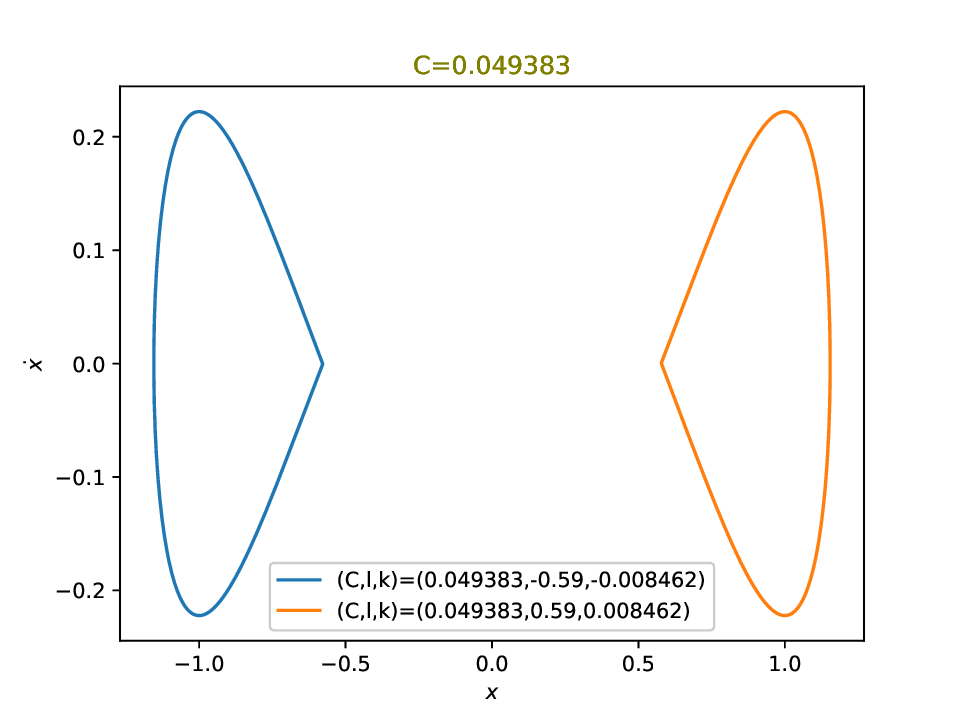}} 
\subcaptionbox{\raggedright\label{cap37}}{\includegraphics[scale=.5]{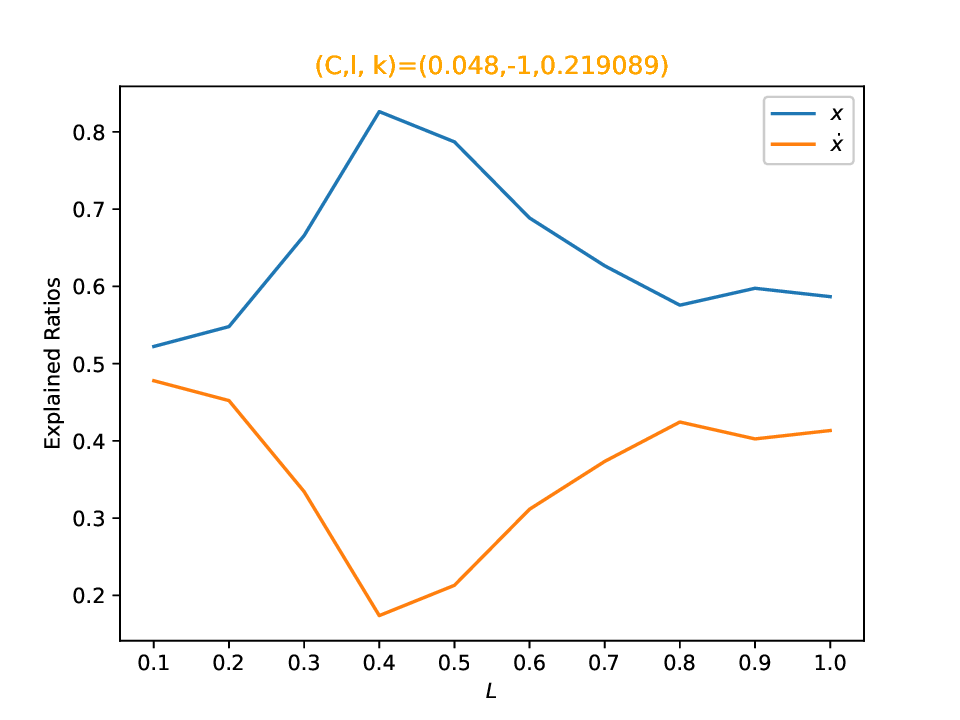}}
\subcaptionbox{\raggedright\label{cap38}}{\includegraphics[scale=.5]{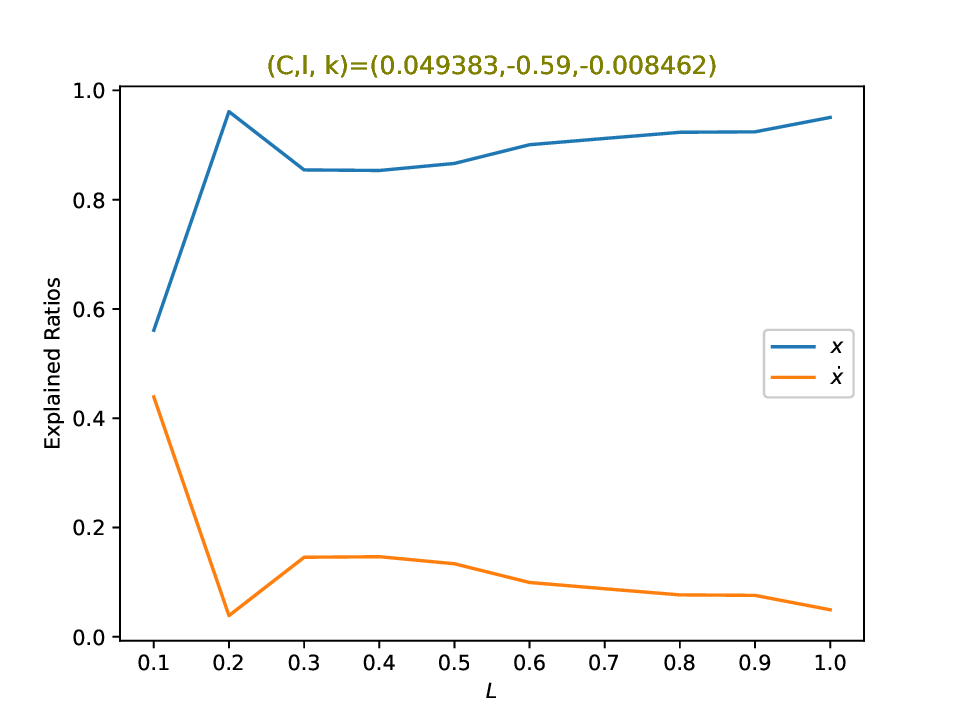}}
\subcaptionbox{\raggedright\label{cap39}}{\includegraphics[scale=.5]{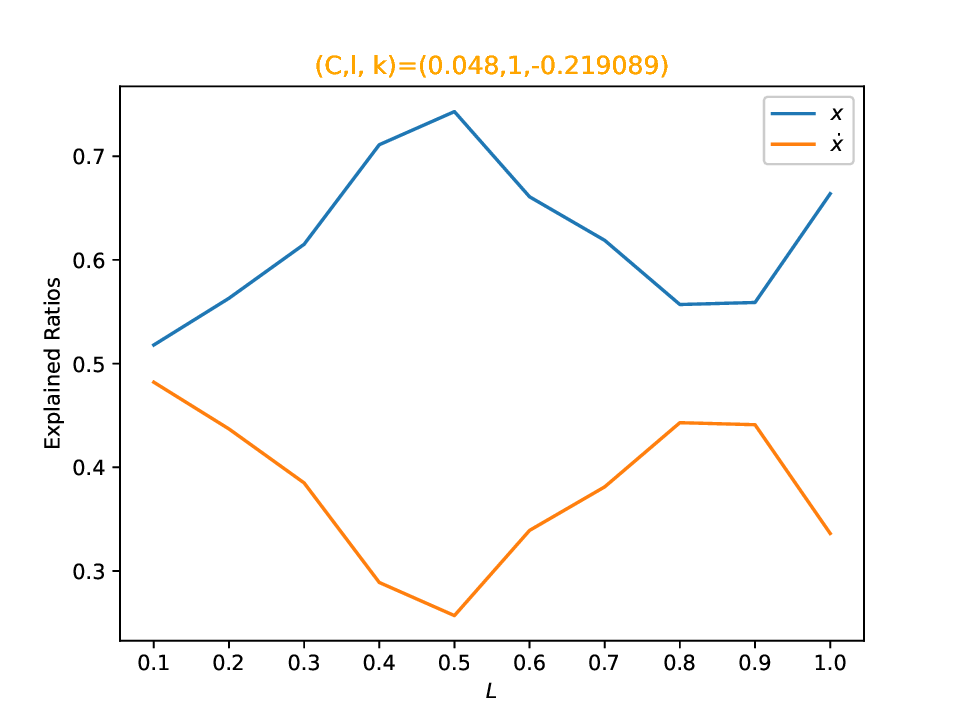}}
\subcaptionbox{\raggedright\label{cap39a}}{\includegraphics[scale=.5]{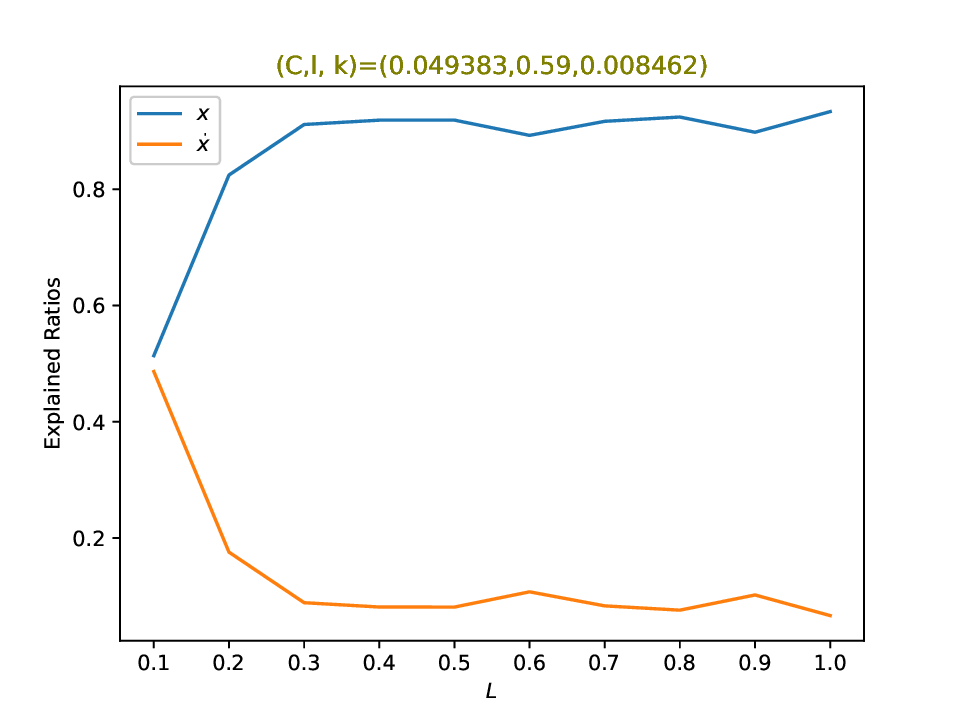}}
\caption{Continued.} 
\end{figure}
\begin{figure}\ContinuedFloat
\centering
\captionsetup[subfigure]{skip=-55mm}  
\subcaptionbox{\raggedright\label{cap39b}}{\includegraphics[scale=.5]{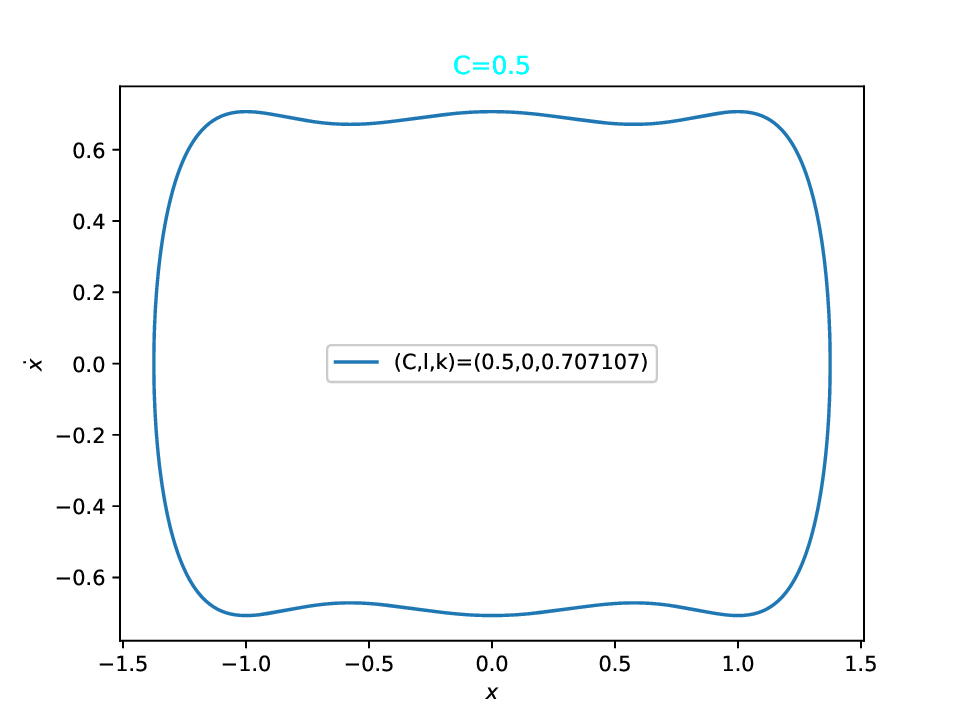}}
\subcaptionbox{\raggedright\label{cap39c}}{\includegraphics[scale=.5]{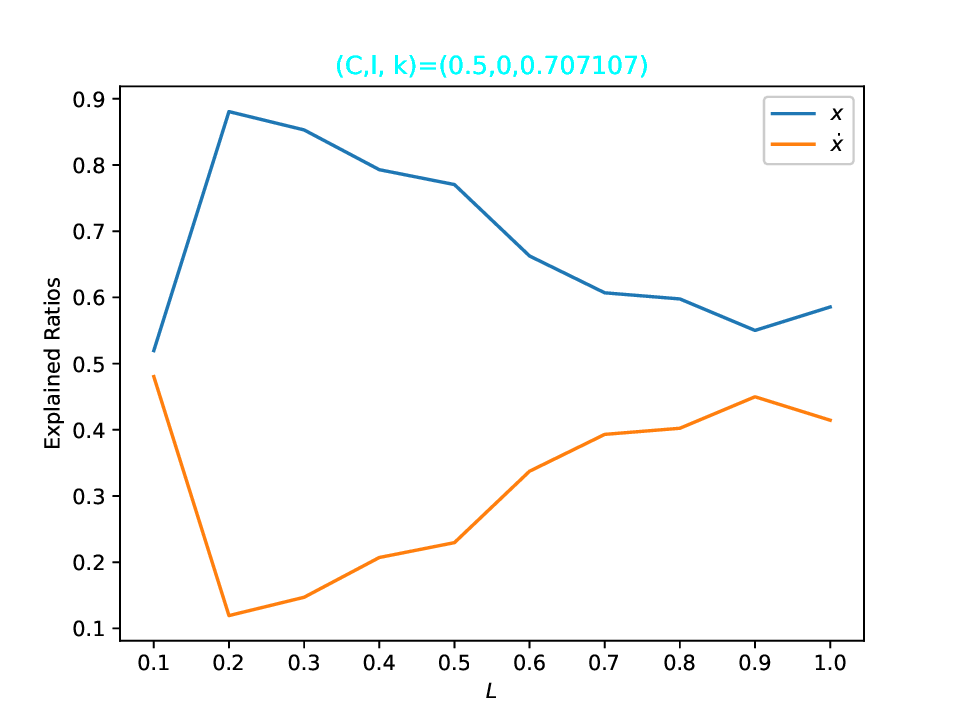}}
\caption{Continued.} 
\end{figure}
  
\subsection{The case $a>1$}

\begin{figure}
    \centering
    \captionsetup[subfigure]{skip=-55mm}
    \subcaptionbox{\raggedright\label{co1}} {\includegraphics[scale=.5]{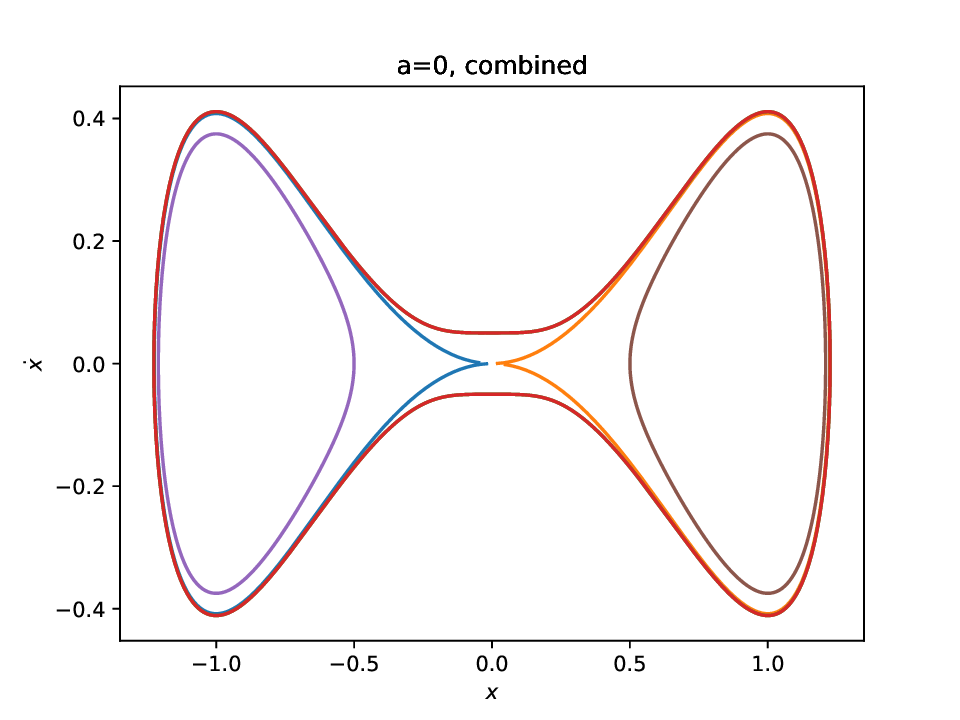}}
    \subcaptionbox{\raggedright\label{co3}} {\includegraphics[scale=.5]{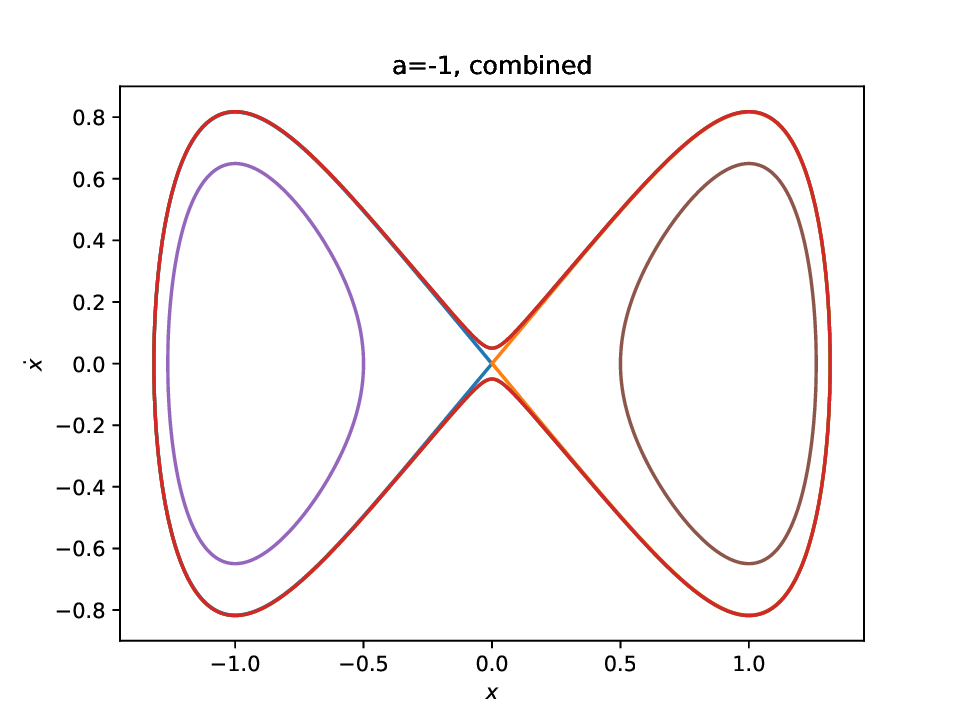}}
    \subcaptionbox{\raggedright\label{co2}} {\includegraphics[scale=.5]{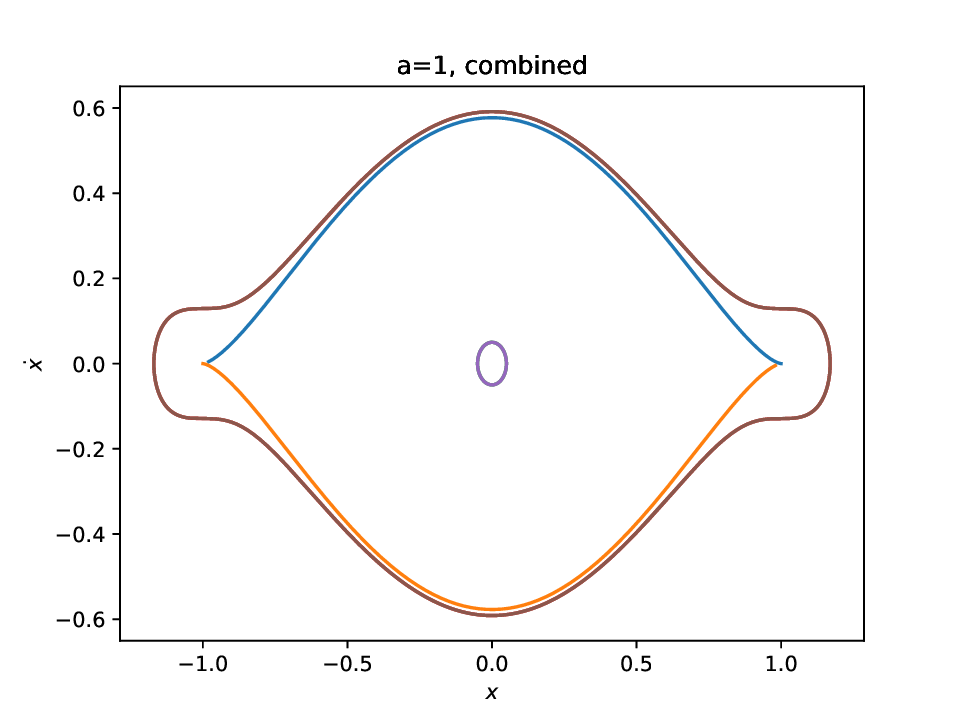}}
    \subcaptionbox{\raggedright\label{co4}} {\includegraphics[scale=.5]{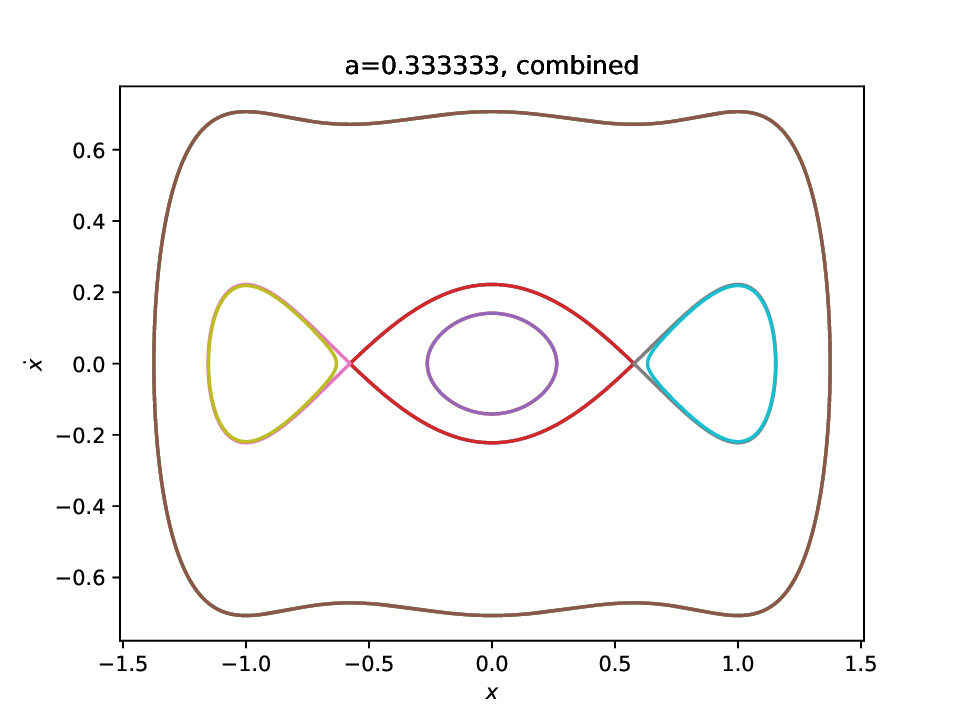}}
    \subcaptionbox{\raggedright\label{co5}} {\includegraphics[scale=.5]{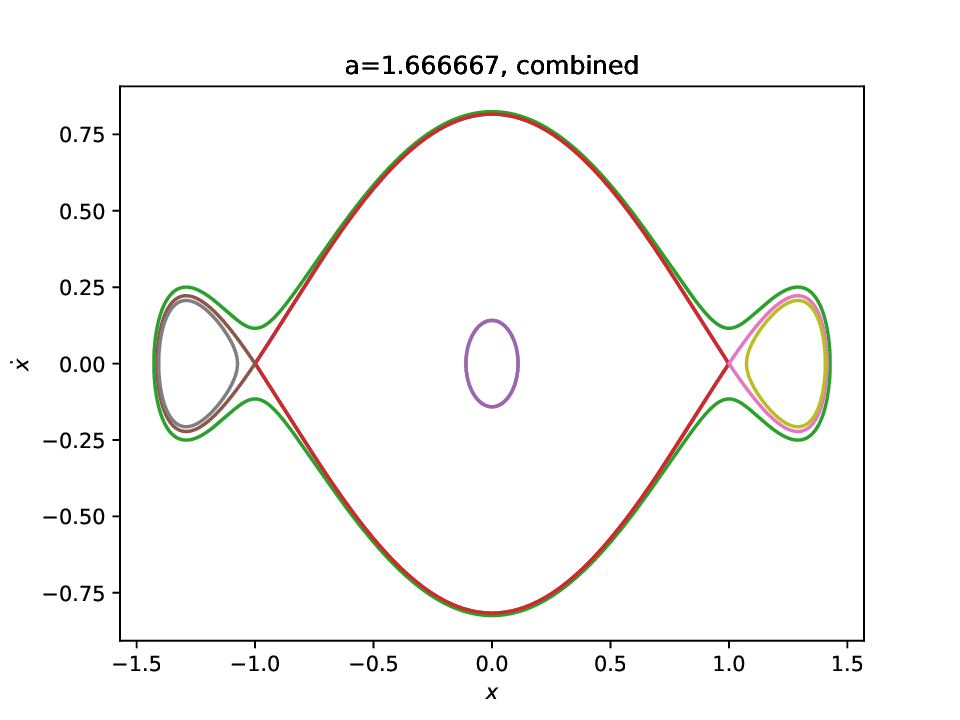}}
    \caption{The phase portraits combined for $a=0, -1, 1, \frac{1}{3}$, and $\frac{5}{3}$. }\label{fc1} 
\end{figure}

Choosing $a> 1$ introduces two more (compared with $a=0$) fixed points. For a typical value of $a=\frac{5}{3}$ in this range, we have
\begin{equation}\label{lm1}
{\dot x}^2=-\frac{1}{3}\,x^2\,(x^4-4x^2+5)+C.
\end{equation}    
There are five fixed points $(0,0), (\pm 1, 0)$, and $(\pm\sqrt{\frac{5}{3}}, 0)$ corresponding to $C= 0, \frac{2}{3}$, and 
$\frac{50}{81}$, respectively. Possible phase portraits and the corresponding explained variance {ratios} are shown in 
Fig. \ref{fd9}. Different trajectories are shown in single plot in Fig. \ref{co5}.
Also, for convenience and comparison, the phase portraits combined are collected in Fig. \ref{fc1} for different values of $a$.

For $0<C<\frac{50}{81}$, there is a trajectory encircling the point $(0,0)$ (plot \ref{cap41}). For $\frac{50}{81}<C<\frac{2}{3}$, there are 
two trajectories each encircling one of the points $(\pm\, \sqrt{\frac{5}{3}}, 0)$ (plot \ref{cap42}). For $C=\frac{2}{3}$ and 
a starting point with {$x=\pm(1-\epsilon)$}, there are trajectories starting there and ending at $x=\mp 1$ (plot \ref{cap45}). 
For $C=\frac{2}{3}$ and a starting point with {$x=\pm(1+\epsilon)$}, there are trajectories starting there, ending at 
$x=\pm 1$, and enclosing one of the points $(\pm\sqrt{\frac{5}{3}},0)$ (plot \ref{cap48}). For $C>\frac{2}{3}$, there is a trajectory enclosing
all five fixed points (plot \ref{cap49}). 

In this case, for $0<C<\frac{50}{81}$ (plot \ref{cap43}) and $\frac{50}{81}<C<\frac{2}{3}$ (plots \ref{cap44} and \ref{cap46}), the difference 
between ratios increases up to $L\sim 0.5$ and decreases afterwards. For $C=\frac{2}{3}$ and $|l|<1$ (plot \ref{cap47}), it has the pattern 
ascending-descending-ascending. For $C=\frac{2}{3}$ and $|l|>1$ (plots \ref{cap49c} and \ref{cap49a}), it increases up to $L\sim 0.4$, then 
remains almost constant, and finally starts to decrease around $L\sim 0.8$. For $C>\frac{2}{3}$ (plot \ref{cap49b}), it increases up to 
$L\sim 0.2$, then remains almost constant up to $L\sim 0.7$, and decreases afterwards. 

\begin{figure}
    \centering
    \captionsetup[subfigure]{skip=-55mm}
    \subcaptionbox{\raggedright\label{cap41}} {\includegraphics[scale=.5]{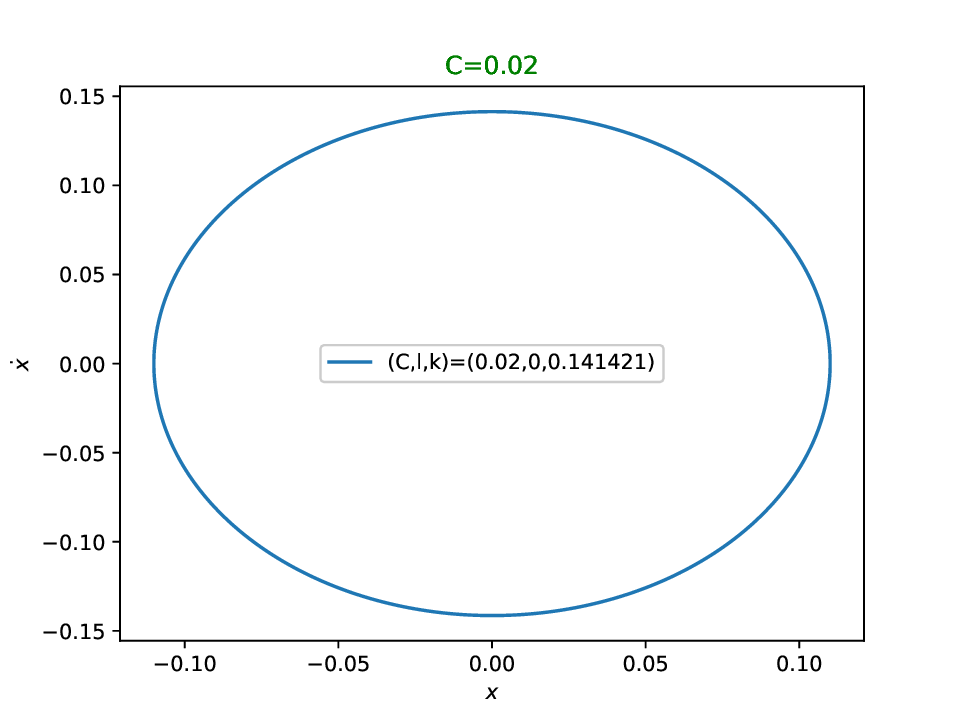}}
    \subcaptionbox{\raggedright\label{cap45}} {\includegraphics[scale=.5]{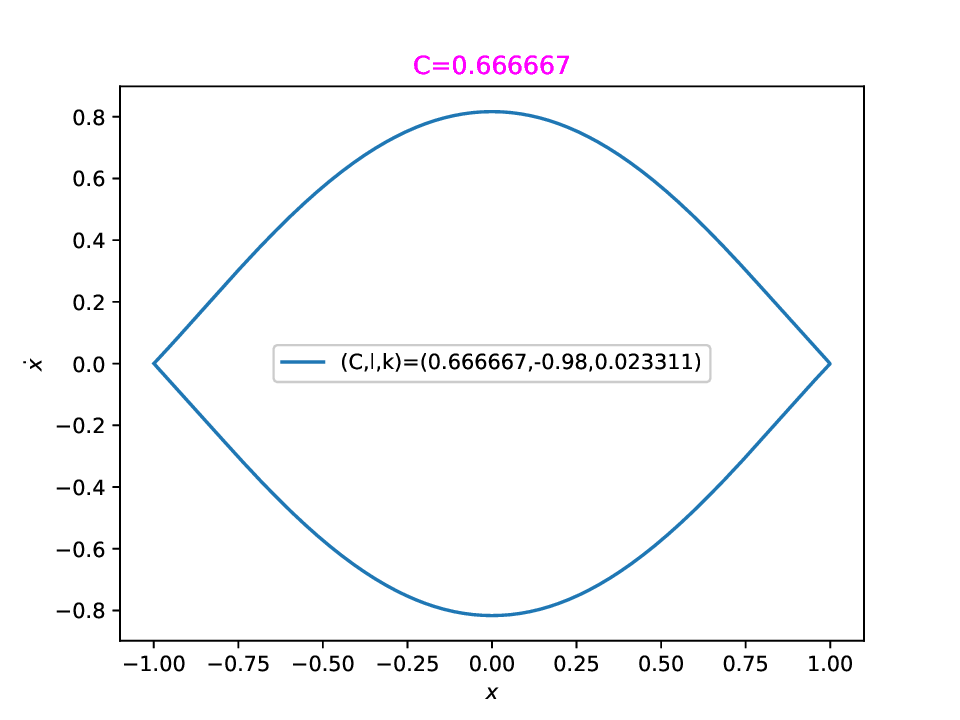}}
    \subcaptionbox{\raggedright\label{cap43}} {\includegraphics[scale=.5]{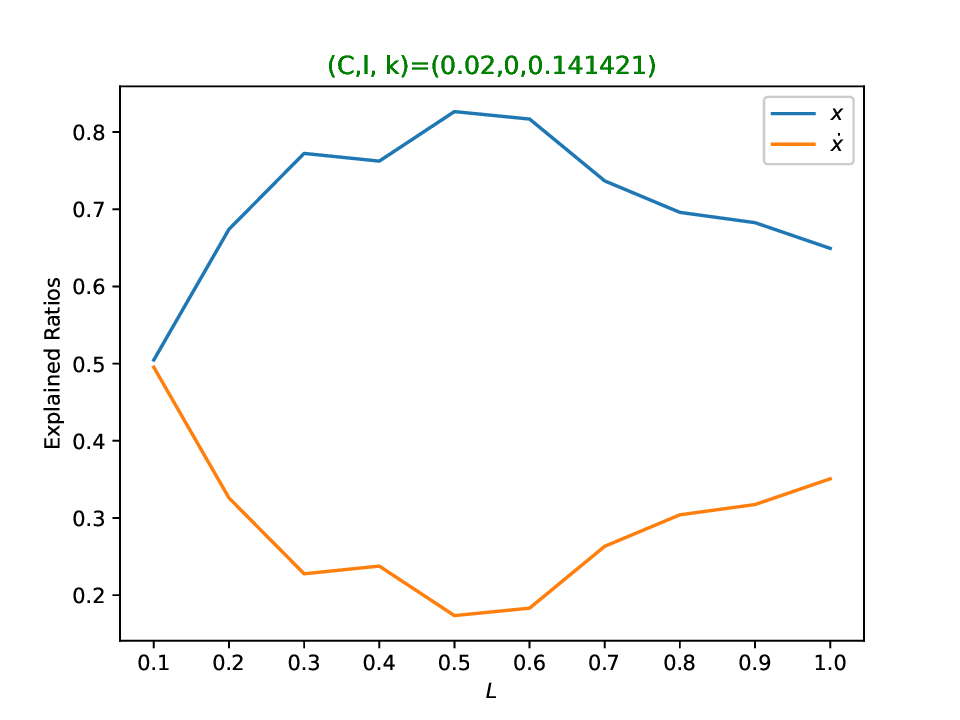}}
    \subcaptionbox{\raggedright\label{cap47}} {\includegraphics[scale=.5]{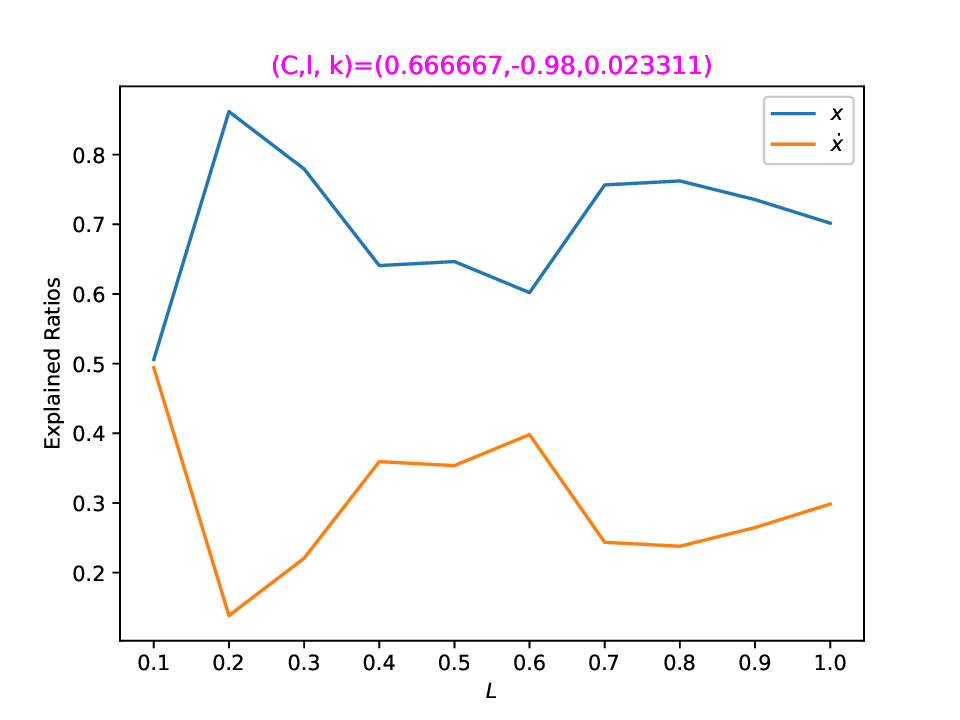}}
    \caption{The phase portraits and explained variance ratios for the unperturbed Duffing equation with $a=\frac{5}{3}$ for 
    $C=0.02$, $l=0$, $k=\sqrt{0.02}\approx 0.141421$ (plots \subref{cap41}, \subref{cap43}), 
    $C=\frac{2}{3}\approx 0.666667$, $l=- 0.98$, $k= \sqrt{-\frac{1}{3}(0.98)^2((0.98)^4-4(0.98)^2+5)+\frac{2}{3}}\approx 0.023311$ 
    (plots \subref{cap45} and \subref{cap47}), $C=0.66$, $l=\pm\sqrt{\frac{5}{3}}\approx\pm 1.290994$,  
    $k=\mp \sqrt{0.66-\frac{50}{81}}\approx\mp 0.206679$ (plots \subref{cap42}, \subref{cap44}, and \subref{cap46}),   
    $C=\frac{2}{3}\approx 0.666667$, $l=\pm 1.02$, $k= \pm\sqrt{-\frac{1}{3}(1.02)^2((1.02)^4-4(1.02)^2+5)+\frac{2}{3}}\approx\pm 0.022849$ 
    (plots \subref{cap48}, \subref{cap49c}, and \subref{cap49a}), and 
    $C=0.68$, $l=0$, $k= \sqrt{0.68}\approx 0.824621$ (plots \subref{cap49} and \subref{cap49b}).}\label{fd9} 
\end{figure}
\begin{figure}\ContinuedFloat
    \centering
    \captionsetup[subfigure]{skip=-55mm}
    \subcaptionbox{\raggedright\label{cap42}} {\includegraphics[scale=.5]{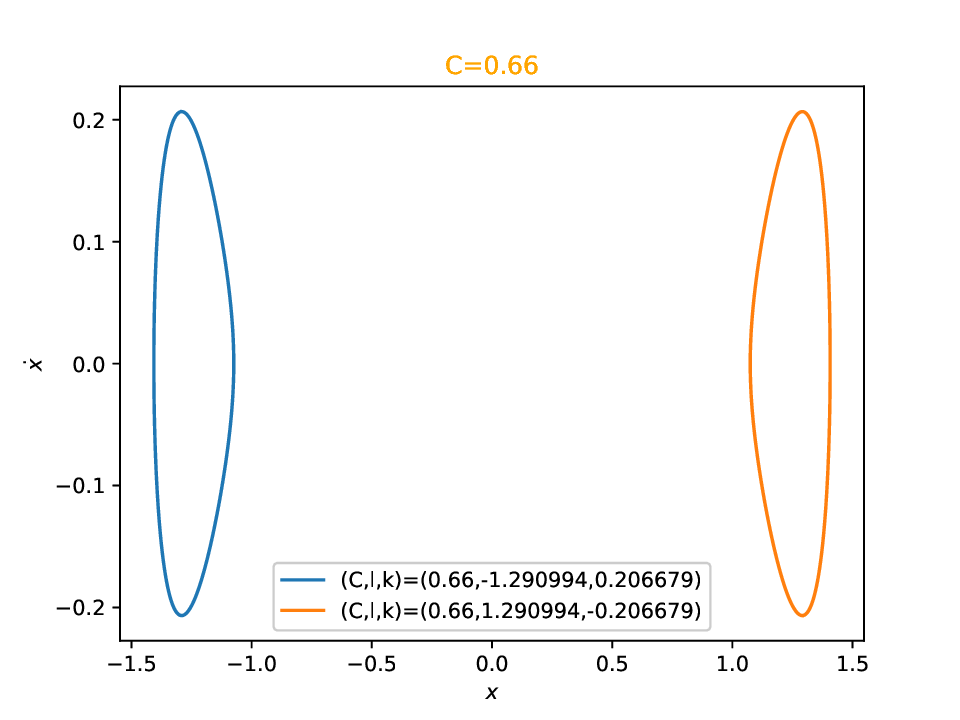}}
    \subcaptionbox{\raggedright\label{cap48}} {\includegraphics[scale=.5]{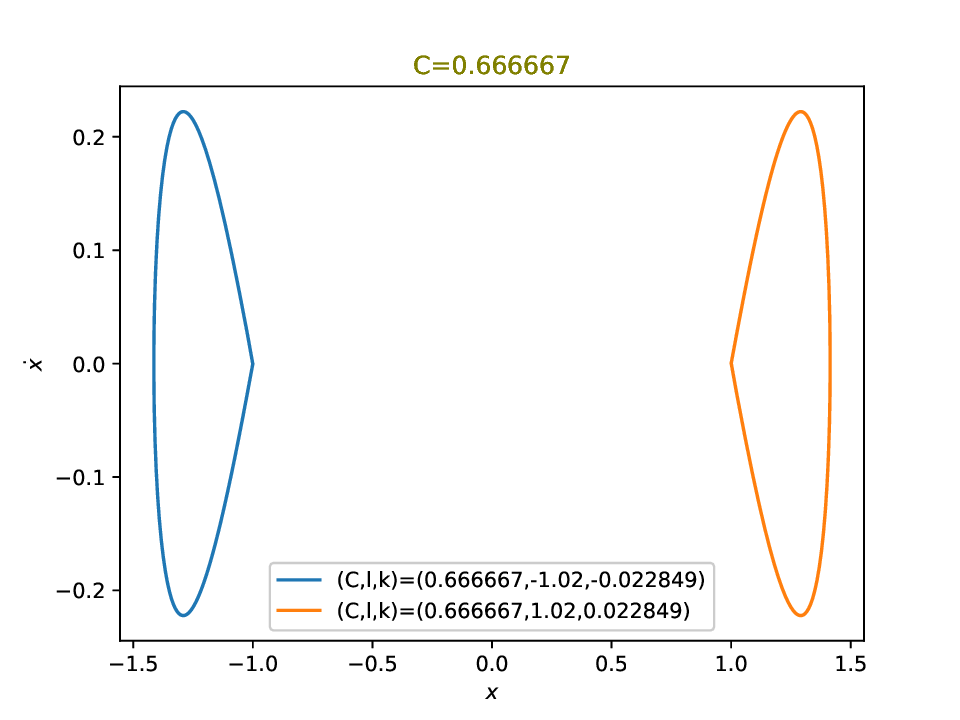}}
    \subcaptionbox{\raggedright\label{cap44}} {\includegraphics[scale=.5]{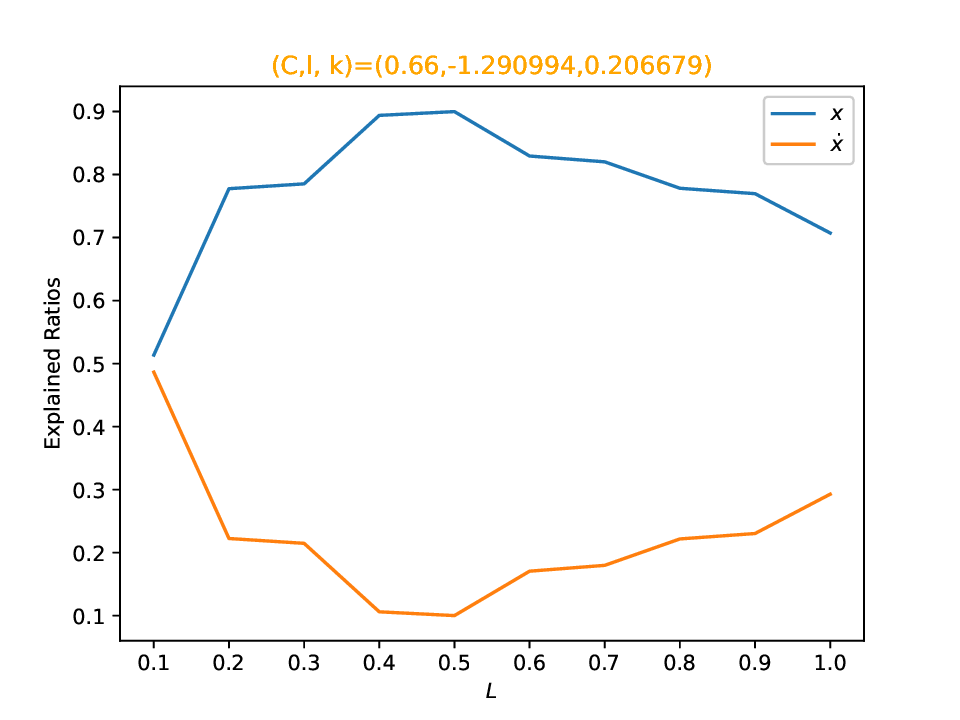}}
    \subcaptionbox{\raggedright\label{cap49c}}{\includegraphics[scale=.5]{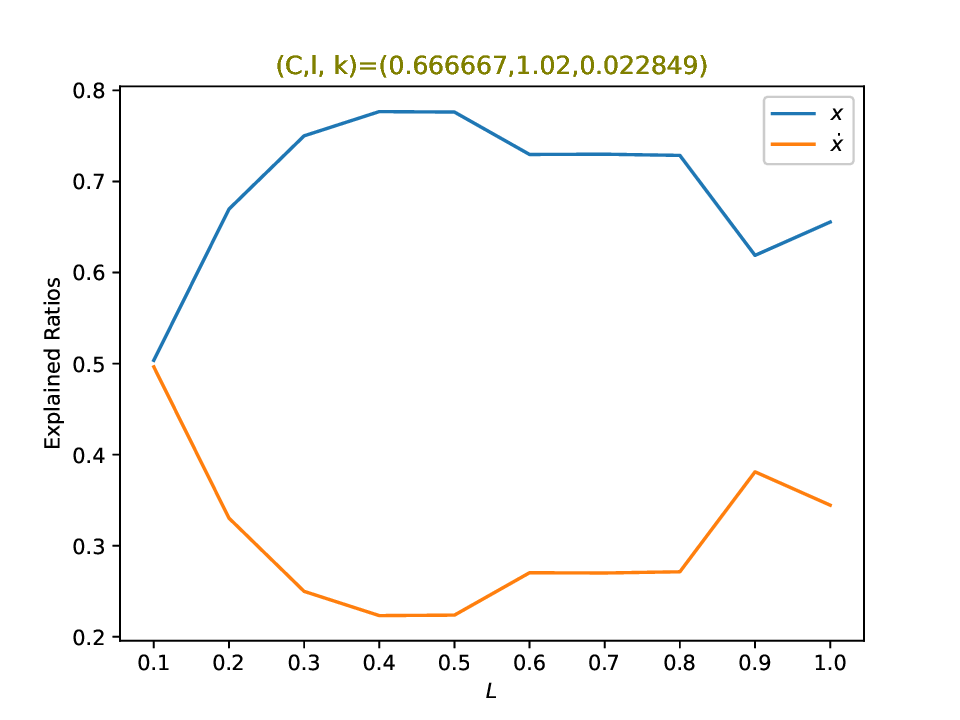}}
    \subcaptionbox{\raggedright\label{cap46}} {\includegraphics[scale=.5]{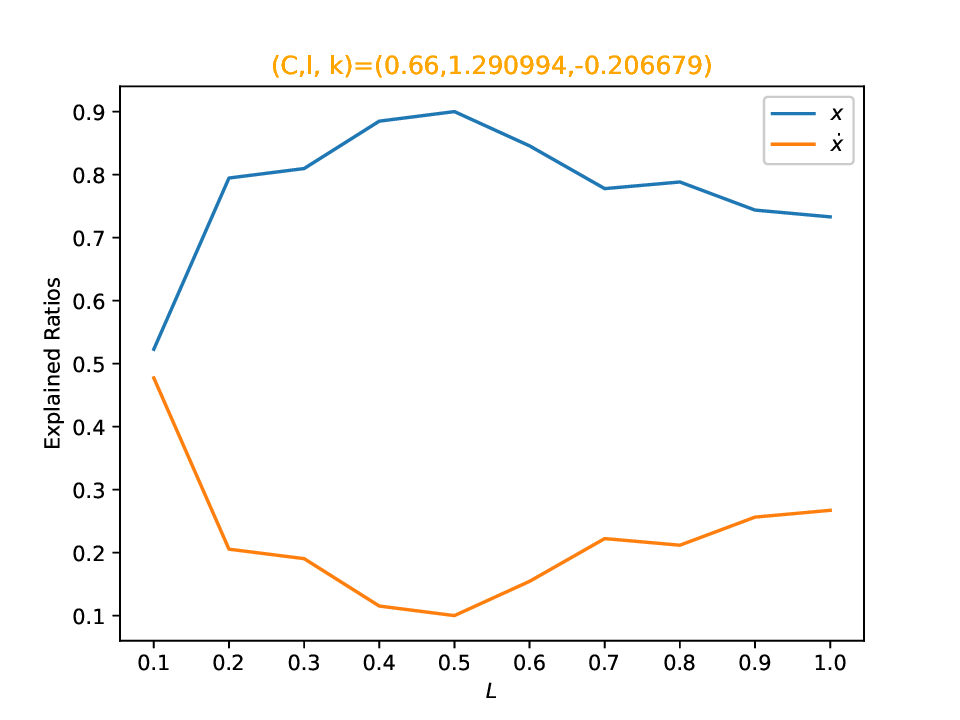}}
    \subcaptionbox{\raggedright\label{cap49a}} {\includegraphics[scale=.5]{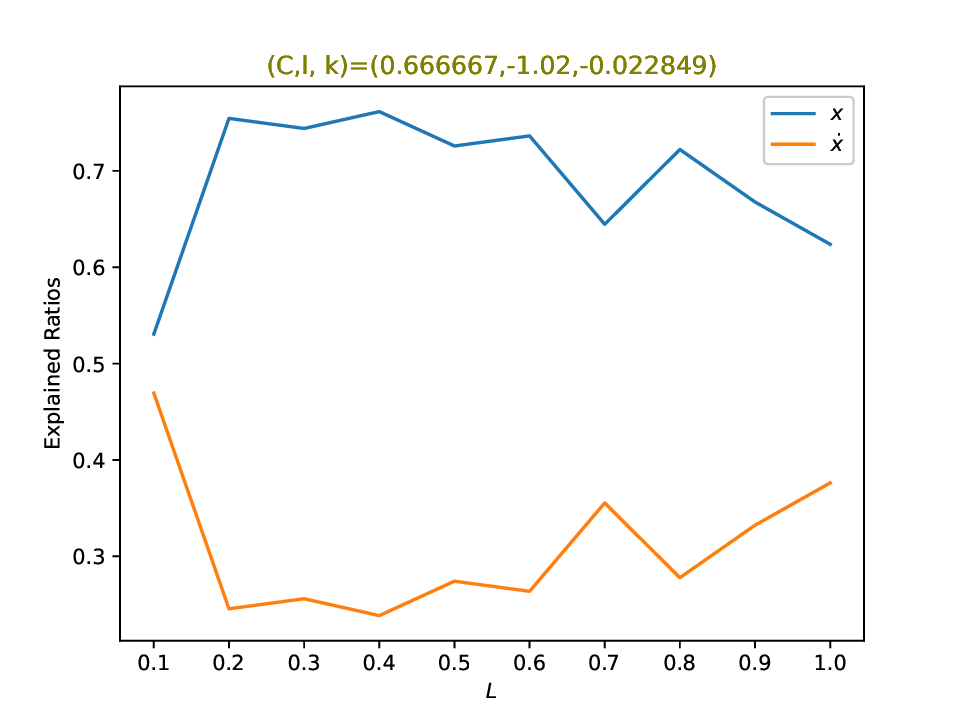}}
    \caption{Continued. } 
\end{figure}
\begin{figure}\ContinuedFloat
    \centering
    \captionsetup[subfigure]{skip=-55mm}
    \subcaptionbox{\raggedright\label{cap49}} {\includegraphics[scale=.5]{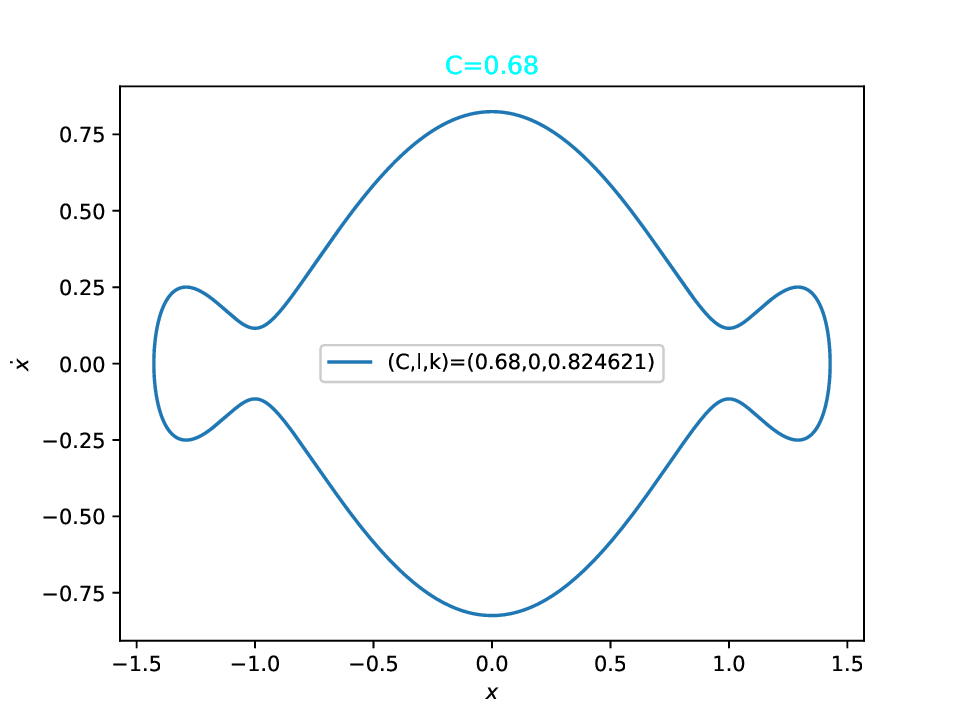}}
    \subcaptionbox{\raggedright\label{cap49b}} {\includegraphics[scale=.5]{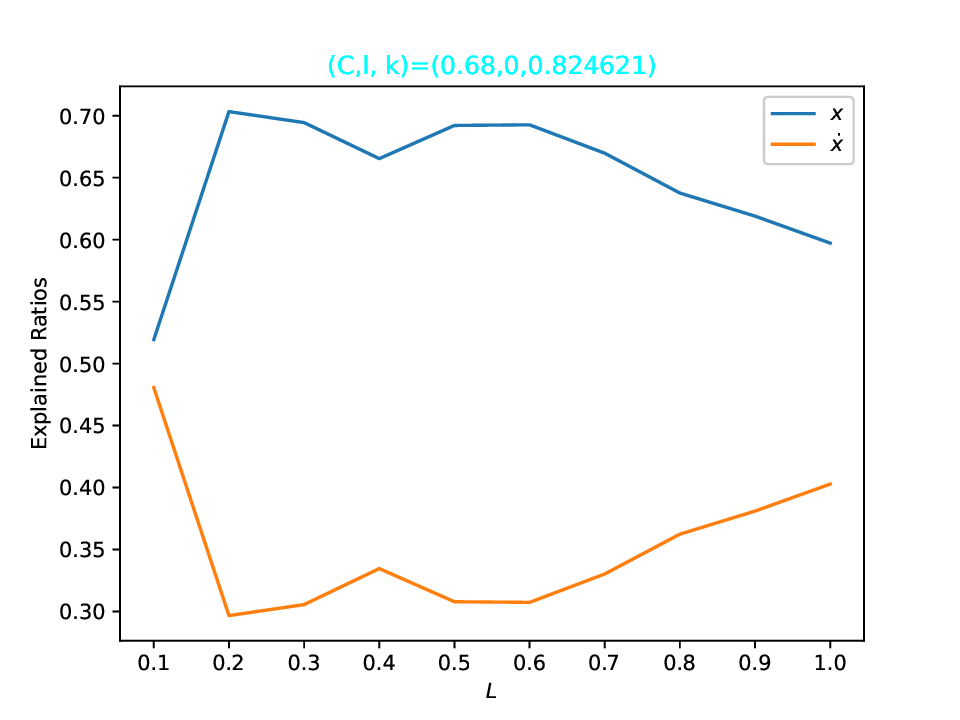}}
    \caption{Continued. } 
\end{figure}

\section{The magnetic reversal}\label{magnetic}

Charged particles in a static magnetic reversal both with and without a shear component have been of interest in certain systems.
The Earth's geo-tail and solar flares are examples of such systems \cite{1992JGR....9715011C}. The effect of the shear component on the 
invariant tori (see e.g., appendix 8 of \cite{arnold1989mathematical} for a description of this concept) of the regular orbits has been studied 
in \cite{1991AdSpR..11i.177B,2002ITPS...30...18Y, YNNERMAN2000217, 2001JPlPh..65..331T}. The intermediate case where the system is not 
integrable nor chaotic has been considered in \cite{2022MNRAS.511.2218L} using the techniques of machine learning.

The magnetic field in the Cartesian coordinates ${\mathbf x}=(x,y,z)$ is given by
\begin{equation}\label{m1}
{\mathbf B}=B_0(h(z),a,b),
\end{equation}
in which $B_0, a, b$ are constants. Two widely used form of $h(z)$ in plasma physics literature are 
\begin{equation}\label{m2}
h(z) = \tanh(z)
\end{equation}
which is called the Harris magnetic field \cite{1962NCim...23..115H}, and $h(z)=z$, which is known as the parabolic field; see 
\cite{2001JPlPh..65..331T} and reference therein. The Harris magnetic field can be produced by a current density which satisfies the
collisionless Vlasov equations; see e.g., \cite{Essen2016}. A force-free generalization of the Harris field has been introduced in
\cite{2009PhRvL.102m5003H} which has non-constant shear components. 

Working in the Coulomb gauge, we choose the following vector potential
\begin{equation}\label{m3}
{\mathbf A} = B_0(s z, bx-F, r x)
\end{equation}
where $s$ is a constant, $r=s-a$, and $F=\int h(z)dz$. In the absence of electric fields, the Hamiltonian of the system is
\begin{equation}\label{m4}
H = \frac{1}{2m}\left({\mathbf P}-q{\mathbf A}\right)^2
\end{equation}
in which $m, q$ are the mass and electric charge of the particle, respectively and ${\mathbf P}$ is its momentum. After some redefinitions, we 
can recast the Hamiltonian into the following form 
\begin{equation}\label{m5}
{\mathcal H} = \frac{1}{2}\left({\mathbf p}-{\mathbf{\mathcal A}}\right)^2
\end{equation} 
in which ${\mathbf{\mathcal A}}=\frac{1}{B_0}{\mathbf A}$. The Hamilton equations, 
$$\frac{\partial{\mathcal H}}{\partial x^i}=-{\dot p}^i,\,\,\frac{\partial{\mathcal H}}{\partial p^i}={\dot x}^i$$ can now be written as
\begin{equation}\label{m6}
\dot{\mathbf X} = {\mathbf J}{\mathbf X}
\end{equation}
where an overdot means $\frac{d}{dt}$, and ${\mathbf X}=({\mathbf x}, \dot{\mathbf x})$ is the state vector of the particle. The explicit form 
of ${\mathbf J}$ is given by  
\begin{equation*}
\setlength\arraycolsep{6pt}
{\mathbf J} =
\begin{pmatrix}
0 & 0 & 0 & 1 & 0 & 0\\
0 & 0 & 0 & 0 & 1 & 0\\
0 & 0 & 0 & 0 & 0 & 1\\
0 & 0 & 0 & 0 & b & -a\\
0 & 0 & 0 & -b & 0 & h\\
0 & 0 & 0 & a & -h & 0
\end{pmatrix}
\end{equation*}

From the equations of motion Eq.(\ref{m6}), we have
\begin{eqnarray}
{\dot x}(t)&=&b\,y(t)-a\,z(t)+l,\label{qe1}\\
{\dot y}(t)&=& F(z)-b\,x(t),\label{qe2}\\
{\ddot z}(t)&=&a\,{\dot x}(t)-h(z){\dot y}(t),\label{qe3}
\end{eqnarray}
in which $l$ is a constant. We solve these equations numerically to find ${\mathbf X}(t)$. Following 
\cite{2002ITPS...30...18Y,YNNERMAN2000217}, we choose $(x(0),y(0),z(0))=(k, 0, 0)$ and 
$({\dot x}(0), {\dot y}(0), {\dot z}(0))=(0, {\dot y}_0, {\dot z}_0)$ which upon inserting into Eq. (\ref{qe1}) results in $l=0$. 
Also, by setting $F(0)=0$, we obtain from Eq. (\ref{qe2}) ${\dot y}_0=-b\,k$. 
We also have ${{\dot z}_0}^2=2E-{{\dot y}_0}^2$ where $E$ is the particle energy, which is constant. 

\begin{figure}
    \centering
    \captionsetup[subfigure]{skip=-55mm}
    \subcaptionbox{\raggedright\label{cap51}}{\includegraphics[scale=.5]{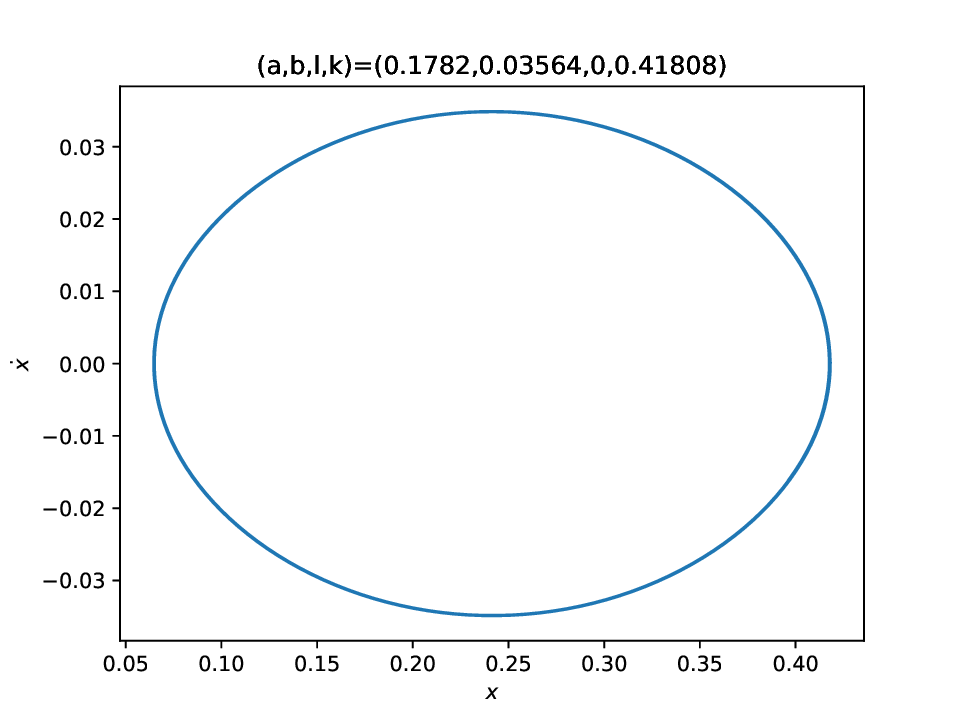}}
    \subcaptionbox{\raggedright\label{cap52}}{\includegraphics[scale=.5]{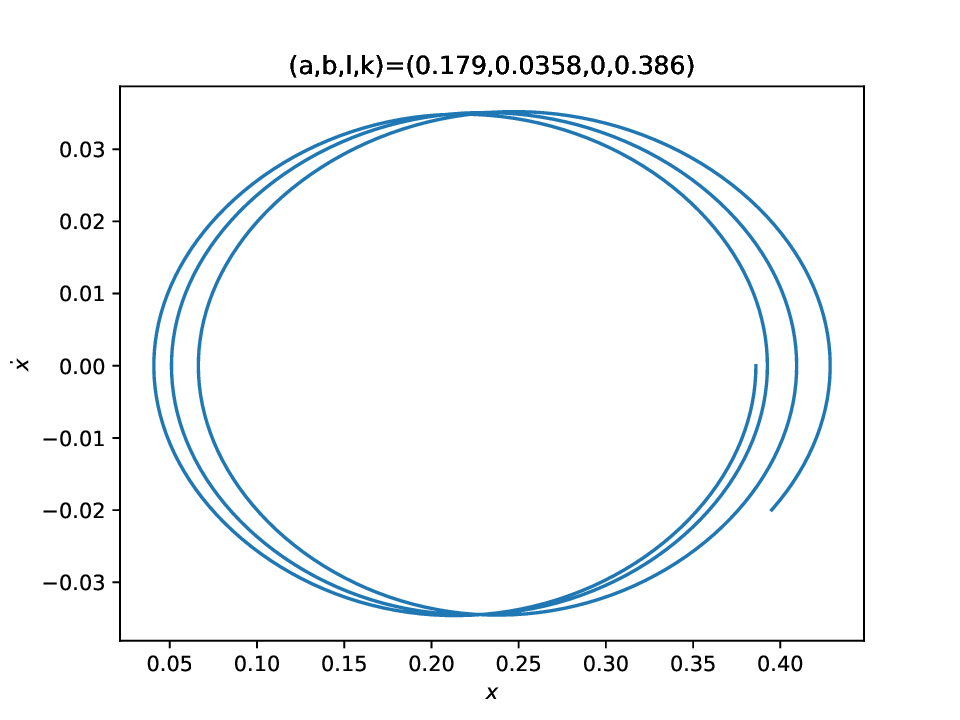}}\\
    \subcaptionbox{\raggedright\label{cap53}}{\includegraphics[scale=.5]{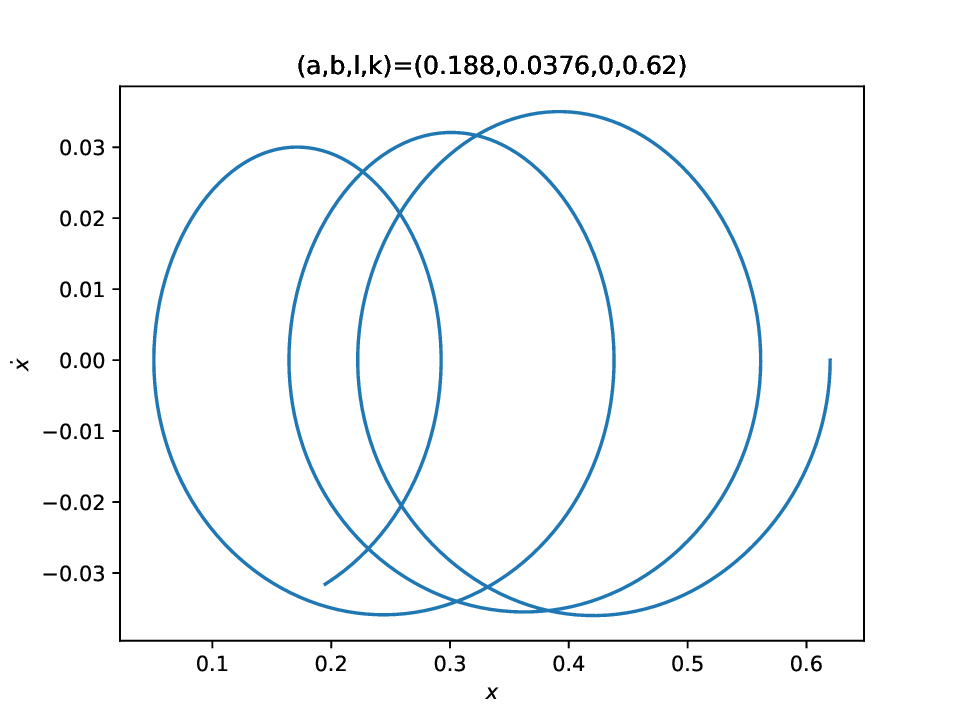}}
    \subcaptionbox{\raggedright\label{cap54}}{\includegraphics[scale=.5]{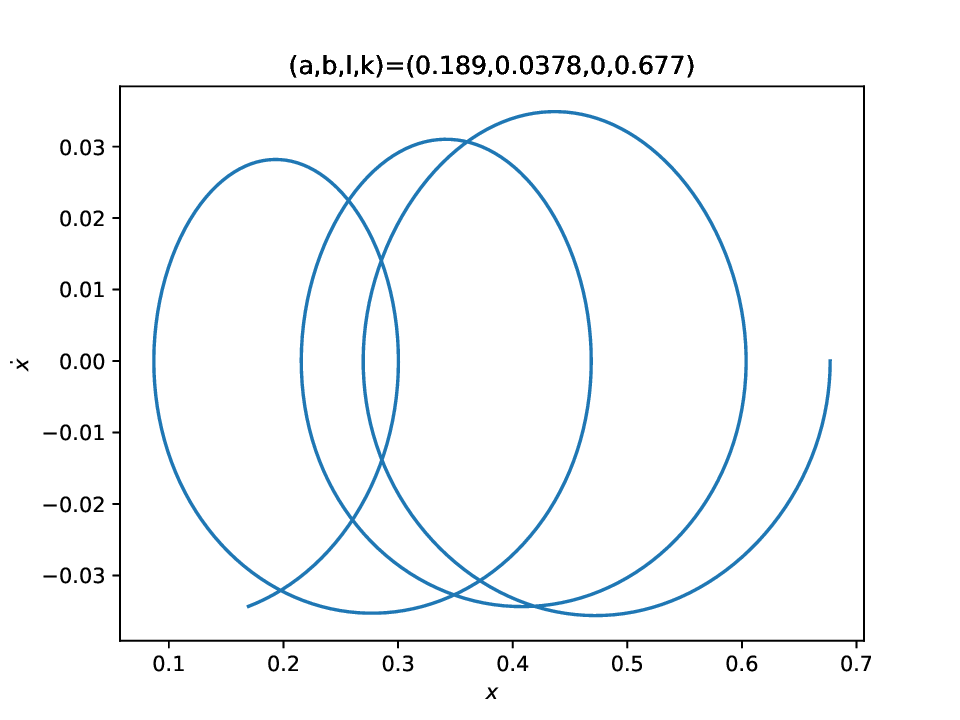}}
    \caption{Some phase portraits in the presence of shear components for $a=0.1782$, {$k=0.41808$} (plot 
    \subref{cap51}), $a=0.1790$, {$k=0.386$} (plot \subref{cap52}), $a=0.188$, {$k=0.620$} 
    (plot \subref{cap53}), $a=0.189$, {$k=0.677$} (plot \subref{cap54}). For all plots, $b=0.2a$ and $E=0.00065$.}
    \label{f2}
\end{figure}

We consider the bifurcation studied in \cite{2002ITPS...30...18Y,YNNERMAN2000217}, which corresponds to $a=0.188$. 
Thus, we take the set of values $({a}, k)=\{(0.1782, 0.41808), (0.179, 0.386), (0.188, 0.620), (0.189, 0.677)\}$, and 
$b=0.2a$, considered in those references.
The phase portraits $x-{\dot x}$, $y-{\dot y}$, and $z-{\dot z}$ and the corresponding explained variance ratios are obtained for the above set
of values. Some of the typical phase portraits are shown in Fig. \ref{f2}. This figure shows change of behavior as $a$ and $k$ are varied.

\begin{figure}
    \centering
    \captionsetup[subfigure]{skip=-55mm}
    \subcaptionbox{\raggedright\label{cap65}}{\includegraphics[scale=.5]{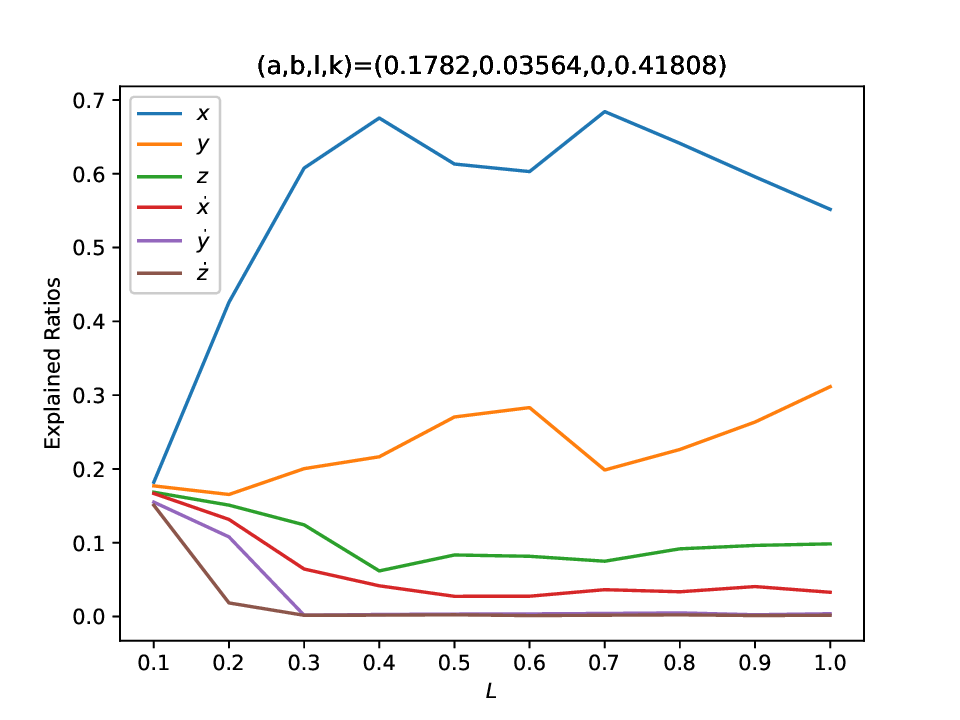}}
    \subcaptionbox{\raggedright\label{cap66}}{\includegraphics[scale=.5]{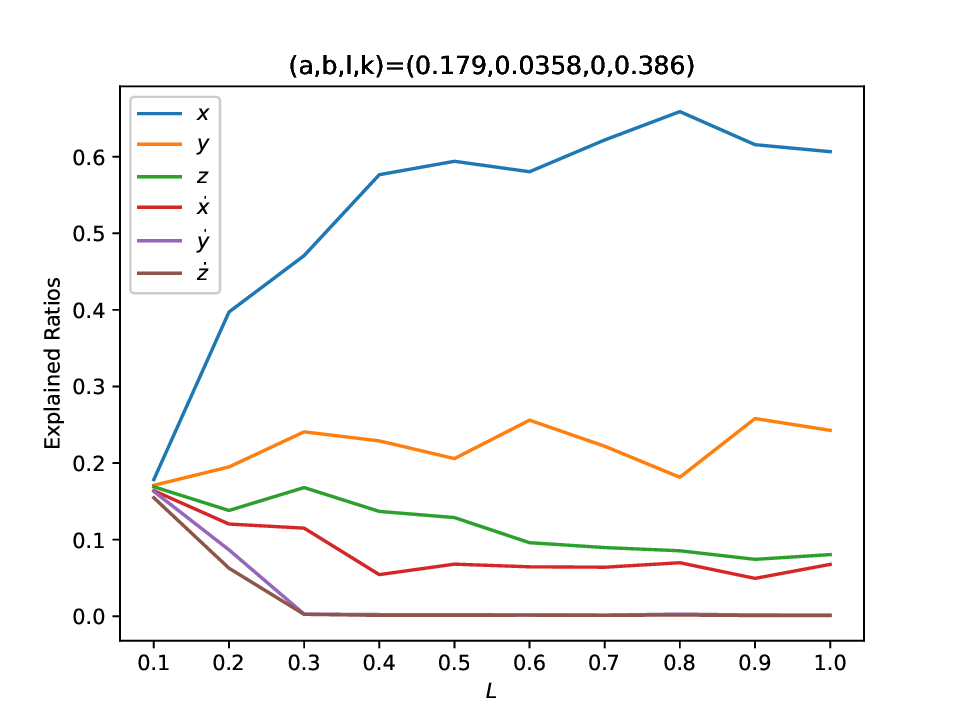}}\\
    \subcaptionbox{\raggedright\label{cap67}}{\includegraphics[scale=.5]{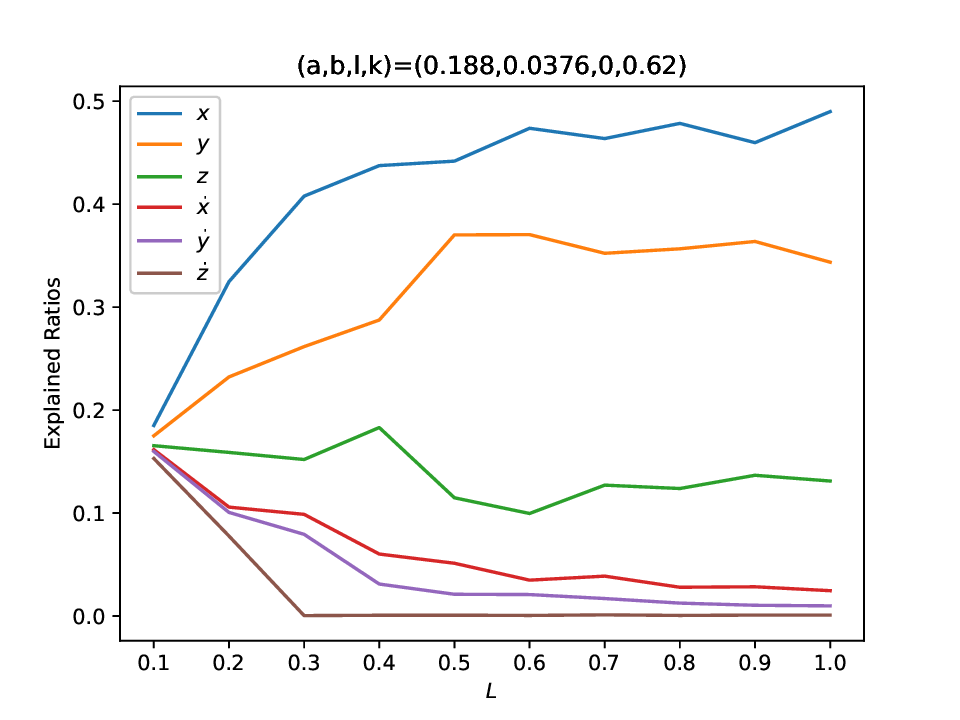}}
    \subcaptionbox{\raggedright\label{cap68}}{\includegraphics[scale=.5]{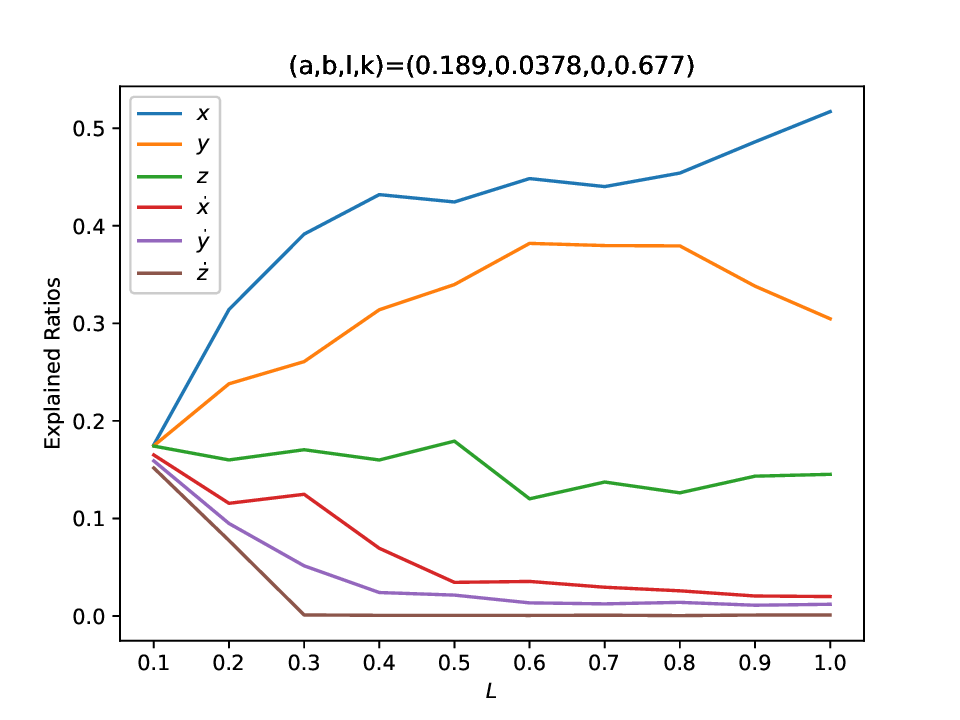}}
    \caption{Explained ratios when the shear components are present for $a=0.1782$, {$k=0.41808$} (plot 
    \subref{cap65}), $a=0.1790$, {$k=0.386$} (plot \subref{cap66}), $a=0.188$, {$k=0.620$} 
    (plot \subref{cap67}), $a=0.189$, {$k=0.677$}  
    (plot \subref{cap68}). For all plots, $b=0.2a$ and $E=0.00065$.}\label{f5} 
\end{figure}

The explained variance ratios are plotted in Fig. \ref{f5} for the above values. The plots clearly show the change of behavior 
when we cross $a=0.188$. 
For $a<0.188$ (plots \ref{cap65} and \ref{cap66}), the number of principal components is $6$ for $L$ up to $0.3$ 
and equals $4$ afterwards. For {$a\geq 0.188$} (plots \ref{cap67} and \ref{cap68}) it is $6$ for $L$ up to $0.3$ and is $5$ afterwards. The 
behavior of the component corresponding to the $y$ direction also changes remarkably. 

{
\section{Results}
For the Duffing equation, a summary of the results is gathered in Tab. \ref{ta1}. In this table $\Delta$ stands for the difference between the
the two explained variance ratios.} 
\begin{table}[h]
    \centering
    \caption{{Behavior of explained variance ratios for the Duffing equation.}}\label{ta1}
    \begin{tabular}{c|c|c}
    \hline 
    $a$  &  $C$  &  $\Delta$ 
    \\ \hline 
                          & $C<0$ & Increases up to $L\sim 0.5$, and decreases afterwards \\\cline{2-3}
                          &  $0$  & Increases up to $L\sim 0.3-0.5$, and decreases afterwards \\\cline{2-3}
    \multirow{-3}{*}{$0$} & $C>0$ & Increases up to $L\sim 0.5$, and almost constant afterwards 
    \\ \hline 
     & $C<0$ & Increases up to $L\sim 0.2-0.5$, and decreases afterwards \\\cline{2-3}
                          & $0$ & Increases up to $L\sim 0.2$, and almost constant afterwards \\\cline{2-3}
     \multirow{-3}{*}{$-1$}                  & $C>0$ & Increases up to $L\sim 0.2$, and decreases afterwards 
    \\ \hline     
                           & $\frac{1}{3}$ & Increases up to $L\sim 0.2$, and fluctuates afterwards \\\cline{2-3}
     \multirow{-2}{*}{$1$}  & $C\neq\frac{1}{3}$ & Increases up to $L\sim 0.2$, and decreases afterwards
    \\ \hline    
                                      & $C<\frac{4}{81}$ & Increases up to $L\sim 0.4-0.5$, and decreases afterwards \\\cline{2-3}
                                      &                                 & For $|l|<\frac{1}{\sqrt 3}$ is ascending-descending-ascending  \\ \cline{3-3}
                                      &\multirow{-2}{*}{$\frac{4}{81}$} & For $|l|>\frac{1}{\sqrt 3}$ increases up to $L\sim 0.2$, and almost constant afterwards  \\ \cline{2-3}
    \multirow{-3}{*}{$\frac{1}{3}$}   & $C>\frac{4}{81}$                & Increases up to $L\sim 0.2$, and decreases afterwards 
    \\ \hline  
                                    & $0<C<\frac{2}{3}$             & Increases up to $L\sim 0.5$, and decreases afterwards \\\cline{2-3}
                                    &                               & For $|l|<1$ is ascending-descending-ascending  \\ \cline{3-3}
                                    &\multirow{-2}{*}{$\frac{2}{3}$} & For $|l|>1$ increases up to $L\sim 0.4$, and almost constant up to $L\sim 0.8$  \\ \cline{2-3}
    \multirow{-3}{*}{$\frac{5}{3}$} & $C>\frac{2}{3}$ &  Increases up to $L\sim 0.2$, constant up to $L\sim 0.7$, and then decreasing
    \\ \hline  
    \end{tabular}
\end{table}

{For the magnetic reversal, the results are summarized in Tab. \ref{ta2} in which $n$ represents the number of
principal components. }

\begin{table}[h]
    \centering
    \caption{{Behavior of explained variance ratios for the magnetic reversal.}}\label{ta2}
    \begin{tabular}{c|c}
    \hline 
    $a$  &  $n$   
    \\ \hline 
    $a<0.188$ & $6$ up to $L\sim 0.3$ and $4$ afterwards 
    \\ \hline 
    $a\geq 0.188$ & $6$ up to $L\sim 0.3$ and $5$ afterwards 
\\ \hline  
    \end{tabular}
\end{table}

\section{Discussion} 
{We proposed an approach to study bifurcations of particle trajectories by deploying deep learning techniques. To examine
this approach}, we studied bifurcations of an anharmonic oscillator described by the unperturbed Duffing equation by deploying the AI Poincar\'e algorithm.
For various values of the parameter $a$ and the initial conditions, we obtained the explained variance ratios corresponding to the possible
phase portraits and investigated their behavior at bifurcations. Changes in the behavior are seen when the parameter or the initial conditions
are varied. These are of the form of changes in the explained ratios growth-decline pattern.

By using the same machinery, we also studied the motion of a charged particle in a magnetic reversal with shear components. A change 
of behavior of the explained ratios is seen in this system {too}. In this case, the change of behavior is quantitative, in 
the form of a change in the number of principal components. This seems to give a more clear description of the bifurcation compared with the 
traditional diagrams. This might be observed in other dynamical systems. In both cases, the explained ratios depend on a noise scale. 

{This approach to study bifurcations of trajectories in dynamical systems is novel. The results provide useful, and in
some cases more clear, signal of bifurcation in addition to the conventional phase portraits and similar tools. In particular, since this 
approach essentially makes use of the principal components analysis, it would be useful in higher dimensional systems where conventional 
tools are more difficult to use, namely phase portraits need to be projected on lower dimensional surfaces. }  

Even though we performed the computations in the frameworks of anharmonic oscillator and magnetic reversal, the 
machinery provided above is not limited to them and can be applied to other dynamical systems accommodating bifurcations. In fact, applications
of the technique to these two prototype dynamical systems shows that the technique works well and is reliable.
It would be interesting to use this technique to study bifurcations in other systems. It would also be interesting to deploy
other algorithms recently proposed for detection of conservation laws to study bifurcations. 

\section*{Acknowledgement}
I would like to thank two anonymous reviewers for their invaluable comments. 


%

\end{document}